# AAO OBSERVER

# Deepest yet
## AAOmega observation probes dark energy

**Starbug Fibre Positioners** | **Time-lapse of the sky over the AAT** | **Tracing the Violent ISM**

# CONTENTS

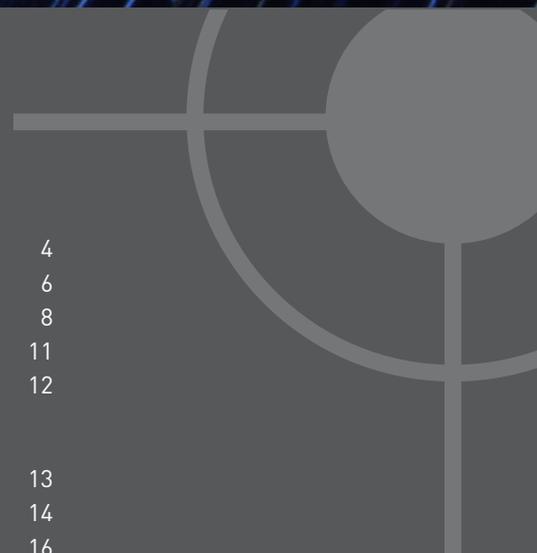

# Director's message

Matthew Colless

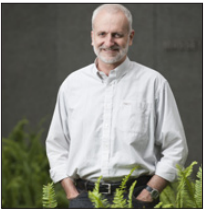

The AAO *Forward Look*[1] is the strategic plan setting out the implementation framework within which the AAO aims to realize the goals and priorities of the *Australian Astronomy Decadal Plan 2006-2015* and the recent *Mid-Term Review of the Decadal Plan*[2]. I have been touring the country over the past couple of months visiting all the major centres in order to present the draft version of the *Forward Look* and get feedback from individuals, institutions, and various advisory committees. This process has been both valuable and encouraging. Although the plan will benefit from further revision in light of the feedback I've received, it's clear that the general shape of the strategy has strong support from the community. Consequently, in order to get a head start on some of the major actions arising from the *Forward Look* recommendations, I am asking AAO staff to begin three important and time-critical initiatives in advance of formal completion of the *Forward Look* process.

**1. Development of instrumentation options for the AAT and UKST.** The Forward Look concludes that high priority should be given to another major instrument for the AAT (see p28, priority #1) and that it is also desirable, if funding can be found, to upgrade the UKST telescope and provide it with a new multi-object spectrograph and positioner system (see p28, priority #4).

For the AAT, the process of identifying the best new instrument will follow the iterative consultative approach successfully used to select HERMES in 2008. That is, the AAO will outline a handful of strawman concepts and put these cases, designs and costs to the community for consideration. These concepts will be weighed and sifted through a workshop process that will eliminate some concepts and perhaps propose others. The AAO will develop this revised set of concepts to the level of costed feasibility studies and then ask the community to evaluate the options and come to a consensus recommendation for the next major AAT instrument. The AAO aims to carry out this process during 2012, noting that HERMES will be commissioned in the first half of 2013.

For the UKST, the science driver is follow-up of the all-sky surveys with ASKAP and SkyMapper. The AAO will examine two potential approaches: a minimal upgrade involving refurbishing the telescope infrastructure and minor improvements to the existing 6dF positioner and spectrograph, and a maximal upgrade including both telescope refurbishment and a new, and fully-automated, 1000-fibre spectrograph and positioner. The trade-off between capital investment and operations costs, as well as the availability of funding, will guide the decision of which (if either) of these options for the UKST the AAO pursues.

**2. Review of AAO's support services for large telescopes.** The AAO currently supports Australian access to large telescopes through operation of the Australian Gemini Office (AusGO), which includes: managing time allocation on Gemini and Magellan through ATAC; assisting with proposal preparation, observation planning and data reduction; providing travel support for observers; taking part in public relations and outreach activities; and participating in instrumentation projects.

The AAO will to carry out a review of the functions and structure of the AusGO during 2012 in order to determine the most effective ways in which it can add value to Australian astronomers' access to Gemini and other offshore large telescopes. The AusGO has been highly effective in the management of Australia's participation in Gemini and Magellan, but it is difficult to determine how much value it is adding to users' observing experience. The AAO is prepared to re-structure the AusGO and potentially provide it with greater resources in order to enhance the service it provides. To this end the AAO needs to consult with the user community and determine what mix of services the AusGO could provide that would improve the research outcomes from access to large telescopes.

**3. Remote observing and AAT operations model.** The AAO aims to phase in remote observing on the AAT, so that by 2015 it is possible to carry out most standard observations on the telescope either classically or remotely. The AAO will also upgrade the AAT's infrastructure and instruments, so that by 2015 the daily burden of repairs and maintenance is reduced by at least 30%.

Remote observing has already been trialled in a limited way from Epping with UCLES, and will be progressively be implemented for other AAT instruments in order of increasing operational complexity. The AAO might initially offer observations from a number of suitably equipped remote locations and, in the longer term, from the astronomer's desk or laptop. This offers potential efficiency savings for users and support astronomers, as well as cost savings for the AAO. Remote observing would be at the user's discretion, and observers would still be welcome to visit the telescope and work in classical mode; the AAO will continue to provide support for visiting observers' travel and living expenses.

Although the AAT manages to keep downtime due to technical problems below 3%, this high level of reliability is presently achieved by continual technical intervention and support. Improving the reliability of the infrastructure and instruments through a program of upgrades and enhancements can substantially reduce the required level of daily support and repairs, and free technical staff at the telescope for strategic project work on new instruments or other telescope improvements.

These are just the first three major initiatives from the *Forward Look*. The plan contains many other improvements, both major and minor, that the AAO intends to implement over the next few years. I encourage you to read the *Forward Look* and see what other enhancements to our facilities and services we will be introducing.

---

1  www.aao.gov.au/AAOForwardLook.html
2  www.science.org.au/natcoms/nc-astronomy/decadalplan.html

# Using 2dF and AAOmega to Harness the Full Power of the Supernova Legacy Survey

Chris Lidman, (AAO and CAASTRO[1]), Vanina Ruhlmann-Kleider (DSM/IRFU/SPP, CEA-Saclay, France) and Karl Glazebrook (Swinburne University and CAASTRO)

The Nobel Prize winning discovery of the accelerating Universe [1,2] was made from observations of a few dozen distant Type Ia supernovae (SNe Ia). The reason for the acceleration is arguably the biggest unsolved mystery in physics. Possible explanations range from vacuum energy to modifications to the theory of General Relativity.

Mapping out in greater detail the expansion history of the Universe is the aim of current SN Ia surveys, as this may offer clues as to what causes the Universe to accelerate. Currently, the largest published individual sample of distant supernovae is the 3-year sample from the Supernova Legacy Survey (SNLS), which contains 252 supernovae extending out z=1.06 [3].

All the supernovae in the SNLS 3-year sample were spectroscopically confirmed. Confirming a Type Ia supernova (SN Ia) at $z \sim 0.7$ from its spectral features typically takes about an hour of integration with an efficient spectrograph on an 8m class telescope. The SNe Ia are observed one at a time, as is it very rare to have more than one SN Ia visible within one week of maximum light at any given time within the fields of view that are available with the current generation of spectrographs on 8m class telescopes. Usually, SNe Ia have to be observed within one week of maximum light, otherwise they are too faint. The observations are, therefore, also time-critical.

This means that many hundreds of hours of time-critical 8m telescope time were required to spectroscopically confirm all 252 SNe Ia in the SNLS 3-year sample. Future surveys, such as the supernova survey of the Dark Energy Survey (DES) [4], will be an order of magnitude larger than SNLS. It therefore seems unlikely that these surveys will be able to spectroscopically confirm all the SNe Ia that they will discover.

## Using AAOmega and 2dF to test a new approach

An alternative yet unproven approach to classifying SNe (and obtaining their redshifts) from real-time spectroscopy is to classify SNe using multi-colour lightcurves and to obtain the redshifts of the host galaxies after the SNe have faded from view. The principle advantages of this technique are i) host galaxy redshifts can be taken at any time, and ii) it usually takes much less time to measure the redshift of a galaxy than to spectroscopically confirm a SN Ia.

In order to test this approach, we recently observed two of the four SNLS fields (D1 and D4) with the 2dF fibre positioner and AAOmega. All 252 SNe Ia in the SNLS 3-year sample were discovered in "real-time", which meant that only part of the lightcurve (usually the rising part) was used to find SNe Ia. Now that the SNLS survey has been completed, it is possible to use the entire data set to find supernovae. Since the entire lightcurve is used, more supernovae (especially fainter and more distant ones) can be found. This has been done in [5]. From the first 3 years of SNLS data, [5] find about 400 SNe of all types per SNLS field. Only about a third of these have a redshift.

Each SNLS field covers 1 square degree, so both the number density of supernova hosts and the area they covered were well matched to the capabilities of the 2dF positioner (400 fibres over a 2 degree diameter field). The median r-band magnitude of the hosts is 23, which is considerably fainter than what is usually attempted with AAOmega. Exposure times were therefore long, up to 60,000 seconds. The two SNLS fields were observed during 4 clear nights during August 2011.

Histograms showing the number of objects that were targeted and the number of objects for which a redshift was obtained are shown in Fig. 1. In total, redshifts were obtained for 398 host galaxies. The completeness fraction is 60%. Most of the galaxies that lack redshifts are fainter than r=23.

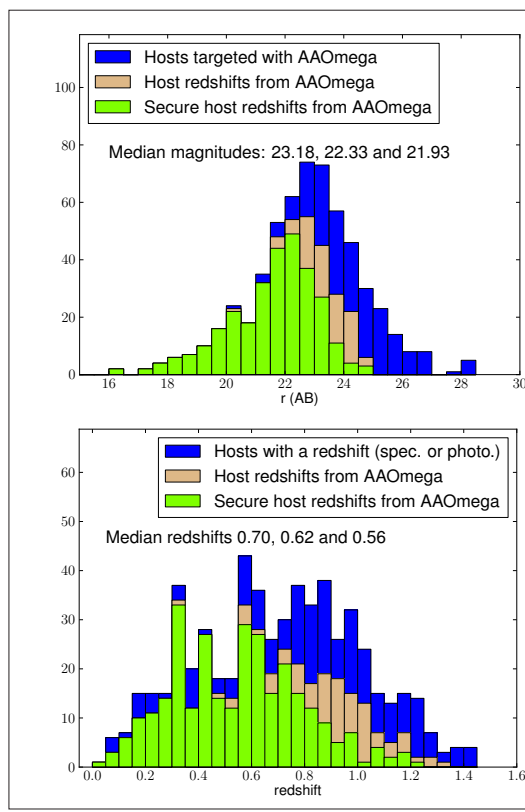

**Figure 1:** Histograms of the number of objects plotted as a function of magnitude (**above**) and redshift (**below**). The blue, green and tan histograms represent all objects targeted with AAOmega, all objects with secure AAOmega redshifts (two or more clearly identified spectral features), and all objects with either a secure AAOmega redshift or a probable one. If the r-band magnitude of the host was unavailable, then the object was not plotted in the upper plot. If a spectroscopic redshift is not available for the lower plot, we use the photometric one. If neither a spectroscopic nor a photometric redshift were available, then the object was not plotted. The median magnitude and redshifts of objects in the blue, tan and green histograms annotate each figure.

---

1  CAASTRO – ARC Centre of Excellence for All-sky Astrophysics

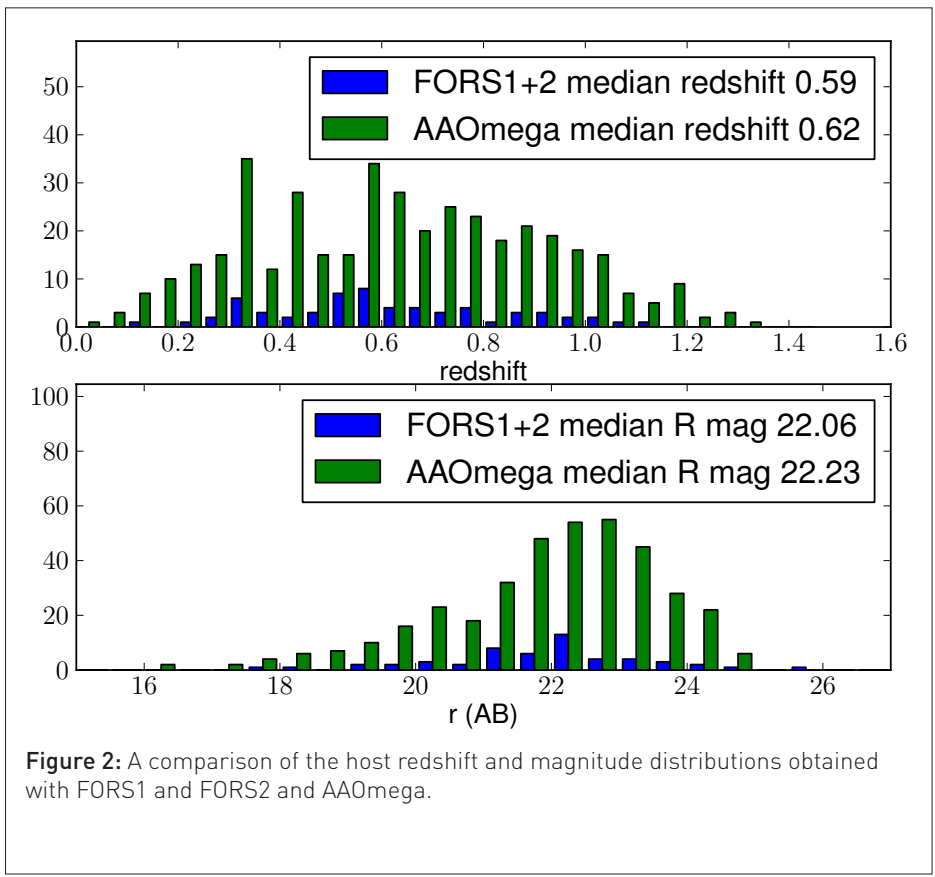

Figure 2: A comparison of the host redshift and magnitude distributions obtained with FORS1 and FORS2 and AAOmega.

## Summary and Future Work

We have demonstrated that we can use 2dF and AAOmega to obtain redshifts for large numbers of galaxies that hosted photometrically classified SNe Ia. The approach is a promising alternative to obtaining a spectrum of every supernova — one at a time — while it is still bright enough to do so. However, before such samples can be used to constrain cosmology, the biases in them need to be examined.

An example of a bias is the relative ease at which redshifts can be obtained for galaxies that are actively forming stars compared to galaxies that don't. If there were no correlation between the properties of SNe Ia and their hosts, then this would not cause a bias. However, there is ample evidence for such correlations. For example, SNe Ia in star-forming galaxies tend to have broader light curves [6], which means that they are also brighter. By preferentially obtaining redshifts for star-forming hosts, one is also preferentially selecting brighter supernova. Uncorrected, this bias might be large enough to significantly change the derived cosmology.

Our observations of the SNLS fields with AAOmega allow us to examine the effect that this and other biases have on the cosmological results, since we can compare the cosmology that was derived for the 3 year SNLS sample with the cosmology that is derived from the sample that uses host galaxy redshifts. The results of this work will have an important bearing on how redshifts are obtained for future SN surveys, such as the SN survey of DES.

The median redshift of galaxies with a redshift from AAOmega is similar to the median redshift of SNe Ia in the 3-year SNLS survey. It is also similar to the median redshift of SNe Ia that will be found in the supernova survey of DES [4].

### A comparison between FORS on the VLT and AAOmega

It is instructive to compare the results we obtain here with the results that we obtained with FORS1 and FORS2 during the real time follow-up of live candidates. During the final two years of the spectroscopic follow-up of candidates with the VLT, the MOS mode of FORS1 and FORS2 was used to observe both live candidates and the host galaxies of other transients that were discovered in earlier years. By the end of the SNLS survey, there were typically three to four host galaxies visible in the FORS 7' x 7' FoV.

In Figure 2, we compare, the magnitude and redshift distributions of objects that were observed with FORS1 and FORS2 with those that were observed with AAOmega. Excluding any sort of renormalisation to account for differences in exposure times, observing efficiency, and target selection, the redshift distributions are broadly similar.

The redshifts from FORS1 and FORS2 came from 66 separate MOS setups. Summed over all setups, the total amount of time spent with FORS1 and FORS2 was 240,000 seconds. For AAOmega, the integration time totalled 92,500 seconds.

At face value, it would seem that AAOmega — a multi-object spectrograph on a 4-metre class telescope - has resulted in many more redshifts than FORS1 and FORS2 — both multi-object spectrographs on 8-metre class telescopes. While this is true, the difference is not as extreme as that suggested by inspecting Figure 2, since the number of targets available for the AAOmega follow-up was about a factor of two larger than the number available for FORS1 and FORS2. The main reason for the difference comes from the difference in the field of view between AAOmega and FORS2. With AAOmega, one can cover the entire 1 sq. degree SNLS field in one shot, which is 60 times the area that can be covered by FORS2. This more than compensates for differences in telescope aperture and image quality.

# Emission Lines in the Near Infrared: Tracing the Violent ISM

Jae-Joon Lee[1], Bon-Chul Koo[2] and the GEMS0[3] team

*The AAO looks for opportunities to exploit its facilities in ways that will provide valuable outcomes for the Australian astronomical community. This includes swapping AAT time in exchange for time on other telescopes and selling AAT time in order to invest in new capabilities (generally new instrumentation). In December 2010 the AAO agreed to sell 15 AAT nights in Semester 2011A to the Korean Astronomy and Space Science Institute (KASI), with the revenue being used to develop new and upgraded AAT instruments. These nights were allocated to a KASI program using IRIS-2 to obtain near-infrared imaging and multi-slit spectroscopy of the interstellar medium around star-forming regions. The following article reports on this research.*

Stars are closely linked to the interstellar medium (ISM) around them. Stars form from the ISM and greatly influence it throughout their lifetimes and as they expire. Jets and outflows from protostars, UV radiation and mass-loss from evolved stars and supernova all heat and excite the ISM and drive turbulent motions within the ISM. Massive stars and novae enhance metal abundances in the ISM, dominating the chemical evolution of Universe. Because these processes are so closely linked, the formation and evolution of stars, particularly massive ones, must be understood in the context of their interaction with the ISM.

Emission lines from the interstellar gas trace the violent interaction between stars and ISM. The H$\alpha$ (6563Å) emission line has been widely used in this regards, and it typically traces diffuse ionized gas. However, due to interstellar extinction, the H$\delta$ emission line can only probe relatively nearby regions in our Galaxy. The interstellar extinction problem is worse for the ISM around massive stars because these regions are often distant and obscured by large columns of gas. Infrared observations suffer less from extinction, and therefore are better suited to studying the ISM near massive stars. Although deeper observations are possible in the far infrared, the near infrared is visible from the ground with the high spatial or spectral resolution critical to probing the kinematics of gas.

The prominent emission lines from the ISM in the near infrared are atomic hydrogen (e.g. Br $\gamma$) and molecular hydrogen (e.g., $H_2$ 2.12 µm). Singly-ionized Iron emission lines (e.g. [Fe II] 1.64 µm) are also important. In particular, $H_2$ lines and [Fe II] lines trace the dense molecular and atomic gas associated with massive stars.

## $H_2$ and [Fe II] emission lines as probes of violent ISM around Stars

Jets and outflows signify star formation. $H_2$ and [Fe II] emission lines probe these jets and outflows effectively (Bally et al., 2007). Both species produce a wealth of lines across the near infrared, allowing us to measure the excitation and kinematics of gas. Collective properties of the jets and outflows in star-forming regions provide estimates on essential parameters of star formation (e.g. star formation efficiency) and shed light on the physics of cloud collapse and star formation in giant molecular clouds. Individually, jets and outflows pin-point the locations of protostars, helping us to understand the formation of stars (Figure 1).

Feedback from massive stars is a critical ingredient of the galactic environment. One key to understanding the evolution of massive stars and their subsequent feedback to their environment is mass loss. Circumstellar shells serve as a fossil record of the mass loss history of the central, evolved stars. [Fe II] and $H_2$ lines have been detected from shells around some evolved massive stars, such as luminous blue variables (LBVs), and are effective probes of their eruptive

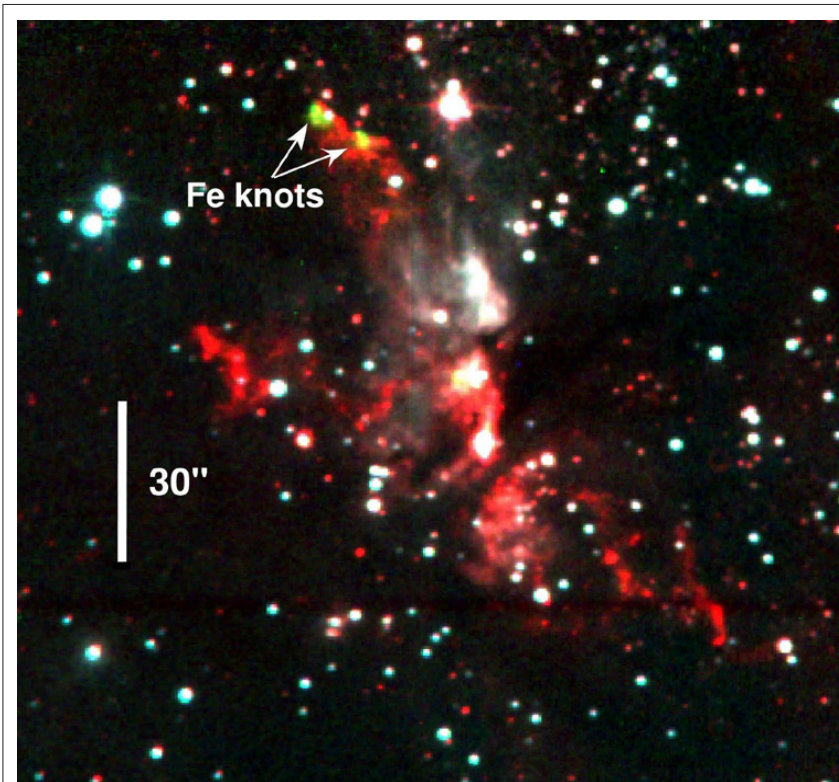

**Figure 1.** A colour-composite image showing the remarkable massive star forming region G35.2N. The prominent $H_2$ outflow with associated [Fe II] knots is clearly visible. Red: $H_2$ 2.12 µm, Green: [Fe II] 1.64 µm, Blue: H band. The H and [Fe II] images are taken with IRIS2. The $H_2$ image is from the UWISH2 survey.

1   Korea Astronomy and Space Science Institute, Korea.
    Email : leejjoon@kasi.re.kr
2   Seoul National University, Korea
3   Galaxy Ecology and Massive Stars at z=0

mass-loss. These emission lines are also found in post-AGB stars and planetary nebulae. They also trace the interaction of supernova remnants with dense interstellar and circumstellar gas well. [Fe II] emission from young core-collapse supernova remnants (SNRs) can provide detailed information on the distribution, dynamics, and chemistry of iron-rich ejecta material. And, the expanding supernova shock waves serve as a time machine to probe the mass loss history of the progenitors, by interacting with dense interstellar clouds or pre-existing dense circumstellar structures and emitting in the near infrared.

### IRIS2 Observations of [Fe II] emission from young core-collapse Supernova remnants

Using IRIS2 on the AAT, our group has been observing star-forming regions, evolved massive stars and supernova remnants. Here we briefly summarize our preliminary results on the near infrared observations of young core-collapse supernova remnants.

Near infrared observations are useful for studies of circumstellar gas in supernova remnants. Of particular interest are young core collapse SNRs with central compact objects and a bright X-ray (of normal abundances) and/or far infrared shell. A good example is G11.2–0.3, with $A_V \sim 13$ mag. We discovered long, clumpy [Fe II] and $H_2$ filaments which we believe are emission from dense presupernova circumstellar wind shocked by a SNR blastwave (Koo et al., 2007). We also discovered faint, knotty features in the interior of the remnant, which turn out to be dense ($\sim 10^3$ cm$^{-3}$) iron ejecta (Moon et al., 2009).

Using IRIS2, we carried out near infrared narrowband imaging and spectroscopic observations of several young core-collapse SNRs in the southern sky. Our preliminary results show that [Fe II] emission is rather common among those SNRs, showing bright filaments and knots (Figure 2). The measured radial velocities of these bright [Fe II] features suggest the emission is not from fast-moving ejecta of the progenitor stars but more likely from circumstellar matter. X-ray emission associated with these [Fe II] emission features also shows no sign of enhanced metal abundances. These observations lead us to speculate that the [Fe II] emission is from dense circumstellar structures that existed before the supernova explosion, which later become shocked by the supernova blast waves. And these pre-exisiting circumstellar structures indicate complicated mass loss history of progenitor stars. The circumstellar structure may arise from the interaction of fast winds of Wolf-Rayet stars with slow, dense winds from a prior red supergiant phase, or it may arise from episodic mass loss associated with interacting binary stars or luminous blue variables. We expect our ongoing studies of these SNRs could provide constraints on the pre-supernova mass loss history of massive stars. Additionally, we searched for faint [Fe II] emission knots with high radial velocities, which could be candidates for fast moving ejecta in young core-collapse SNRs. We have detected [Fe II] knots showing velocities of several hundred km/s, and we are investigating the origin of these features.

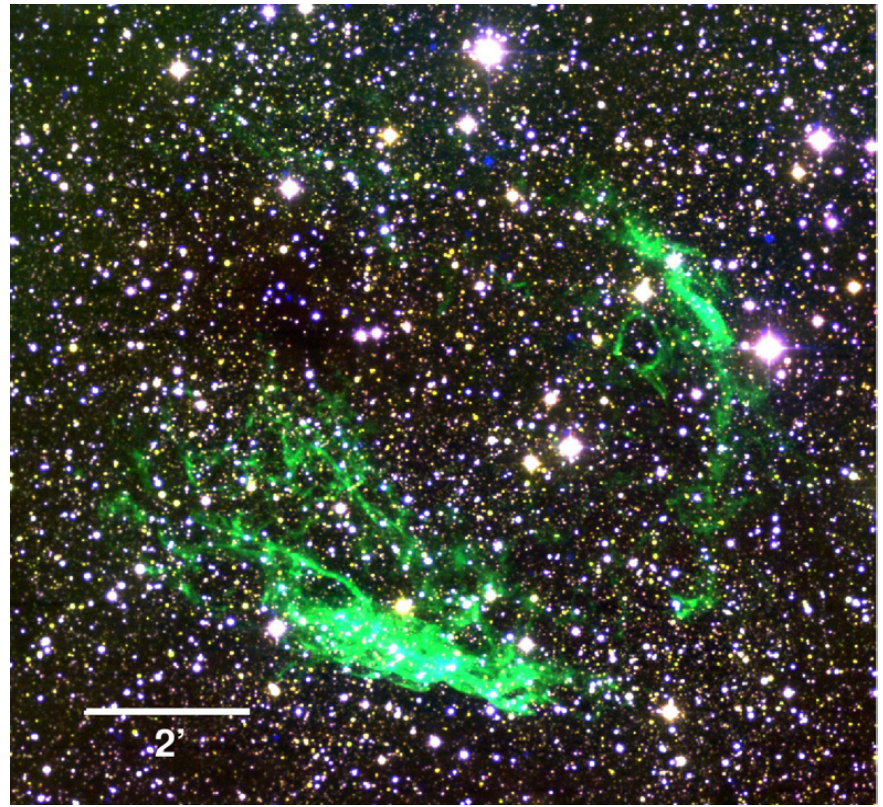

**Figure 2.** A colour-composite image showing the Galactic supernova remnant RCW 103. Red: $H_{cont}$, Green: [Fe II] 1.64 μm, Blue: J band. The $H_{cont}$ and [Fe II] images are taken with IRIS2. The J band image is from the 2MASS survey.

### A Near Infrared Narrowband Wide-field Imaging Survey

Our own Galaxy and its inner workings of star birth and death are still mysterious. Unbiased imaging surveys of the Galactic plane in broadband infrared, such as Glimpse and MIPSGAL, have greatly increased our understanding of our Galaxy. A narrowband imaging survey can complement broadband surveys by focusing on emission line objects. An excellent example of such a narrowband survey is the UKIRT Widefield Infrared Survey for $H_2$ (UWISH2; Froebrich et al., 2011). This survey covered 150 square degrees along the Galactic Plane (10° < l < 65°; -1° < b < +1°) with WFCAM at UKIRT to detect the molecular hydrogen 1-0 S(1) emission line at 2.12 μm and probe a dynamically active component of star formation not covered by broad-band surveys. Our group is planning a similar unbiased survey using WFCAM at UKIRT with a narrowband filter centered at [Fe II] 1.64 μm. We hope our survey, once completed, will complement the UWISH2 survey and other broadband surveys to enable a better understanding of our Galaxy and provide the community with a legacy dataset useful for a wide variety of astrophysical investigations.

# Dancing Starbugs: vacuum adhesion, field rotation and other progress

James Gilbert, Jeroen Heijmans, Michael Goodwin, Stan Miziarski, Rolf Muller, Will Saunders, Alex Bennet, Julia Tims (AAO)

The past year, 2011, was a productive year for the Starbug fibre positioner concept. In fact it hardly seems fair to call it a concept these days. Driven principally by the proposed Many Instrument Fibre System (MANIFEST) facility for the Giant Magellan Telescope, Starbugs are now performing better than ever.

Starbugs are miniature 'walking' robots developed to overcome the limitations of sequential 'pick and place' positioners by moving many optical payloads simultaneously. Starbugs provide fast field configurations of several minutes with ability to operate over large focal planes. A Starbug prototype is shown in **Figure 1**. Their simple design incorporates two piezoceramic tubes to form a pair of concentric 'legs' which, with the application of specific waveforms, can produce a micro-stepping motion in $\pm x$ and $\pm y$ directions. The robots have a clear central aperture for the insertion of fibre bundles or fore optics (see **Figure 2**). For the MANIFEST design, the Starbugs hang upside down beneath a curved glass field plate (see **Figure 3**). This eliminates the need for fibre retractors, but also presents a number of new challenges.

### Defying gravity

Early Starbug prototypes for MANIFEST used a pair of magnets to clamp each Bug to the field plate (see **Figure 4**). This necessitated a very low friction coating on the top side of the plate so that the Starbugs could drag along their respective 'slave magnets'. Not only was it difficult to source a sufficiently wide-band coating with a very low friction coefficient, but also the Starbugs struggled to tow this heavy magnetic 'ball and chain' around, especially up an inclined field plate. A recent advancement has eliminated many of these problems: a vacuum is now used to adhere Starbugs to the field plate, with a thin silicone hose connecting each Bug to an evacuated reservoir (see **Figure 5**). The removal of the magnet assembly has resulted in major improvements to Starbug performance, as well as a reduction of their overall diameter and therefore their minimum spacing. These advantages have so far outshone the disadvantages of using an active vacuum system.

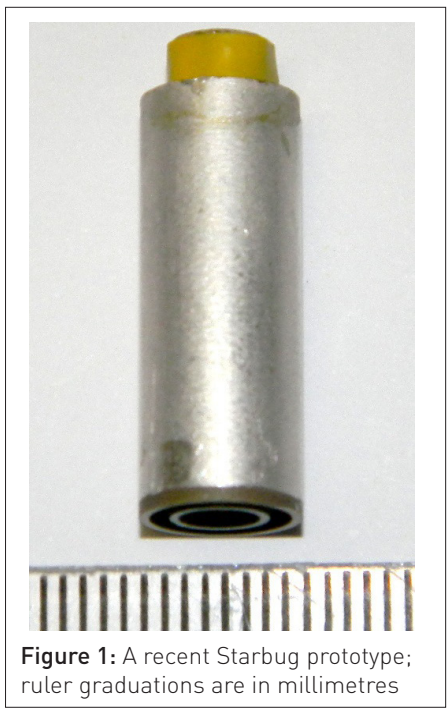

**Figure 1:** A recent Starbug prototype; ruler graduations are in millimetres

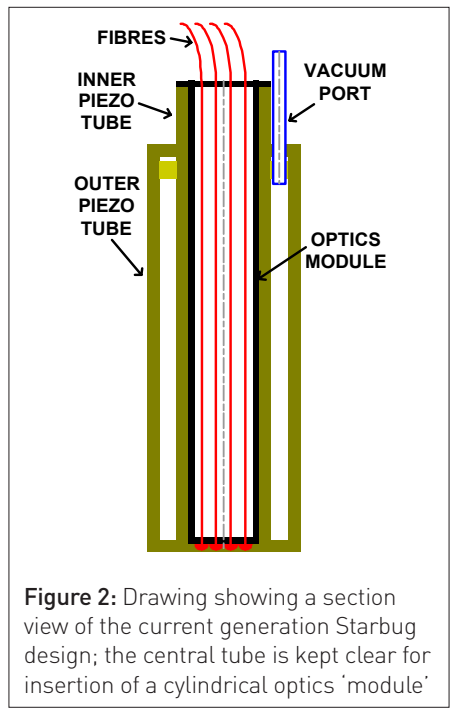

**Figure 2:** Drawing showing a section view of the current generation Starbug design; the central tube is kept clear for insertion of a cylindrical optics 'module'

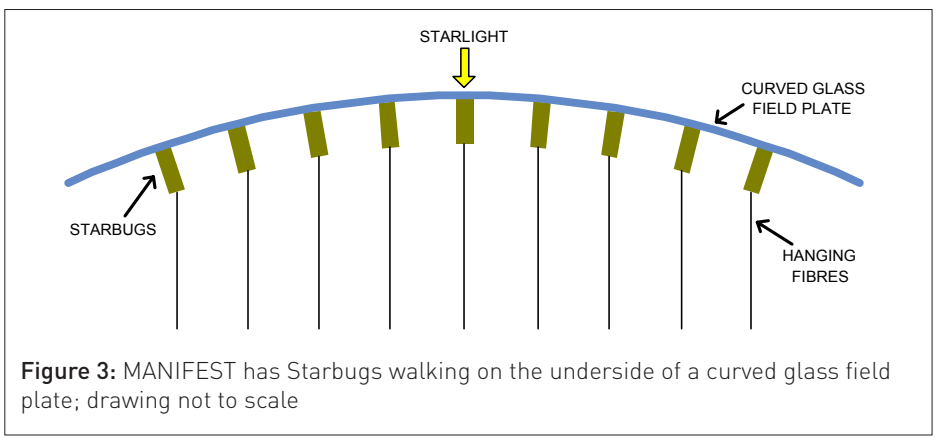

**Figure 3:** MANIFEST has Starbugs walking on the underside of a curved glass field plate; drawing not to scale

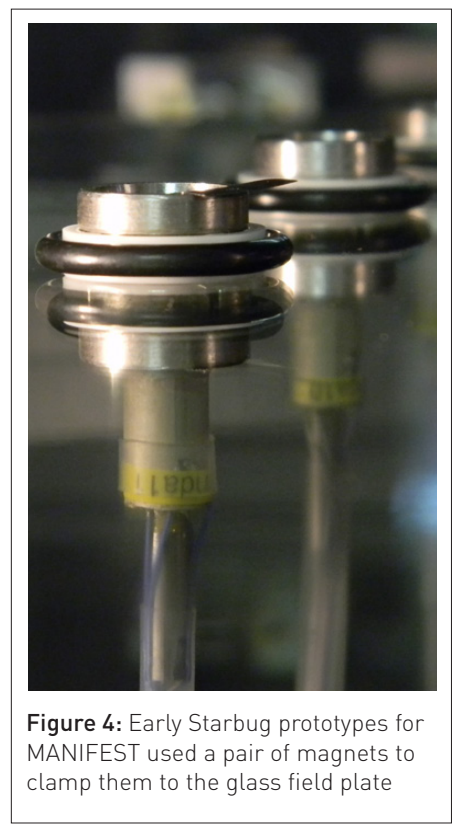

**Figure 4:** Early Starbug prototypes for MANIFEST used a pair of magnets to clamp them to the glass field plate

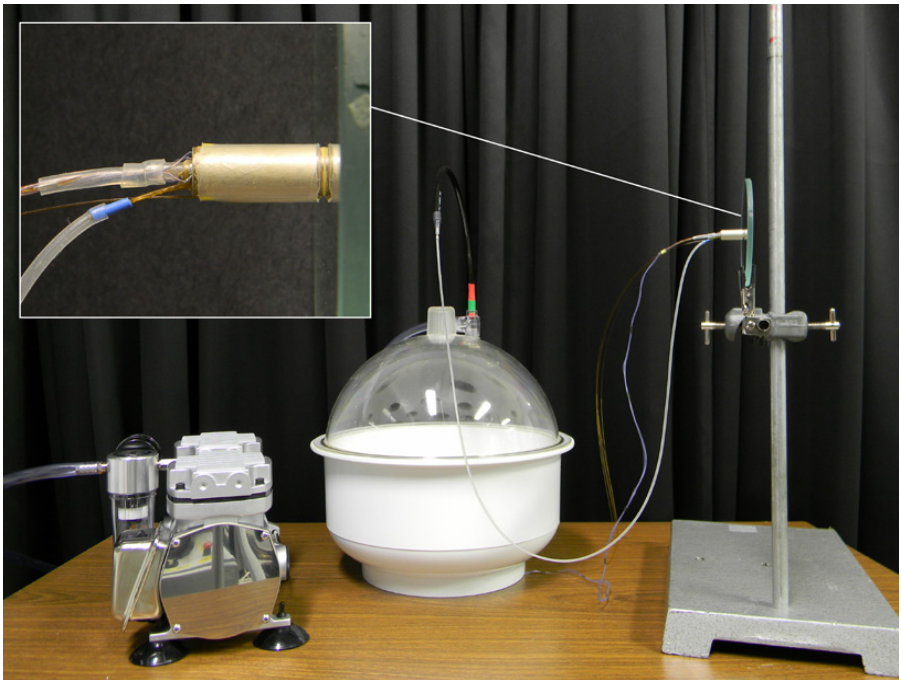

Figure 5: A demonstration of the new vacuum adhesion method, showing a single Starbug on a vertical glass surface

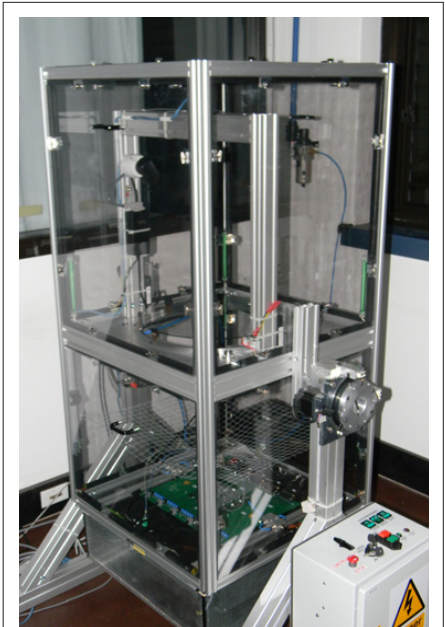

Figure 6: The Starbugs 'test rig' has control electronics and metrology for up to 20 Starbugs; in the centre of the enclosure is an interchangeable 400 mm diameter glass field plate

### Field rotation

An important breakthrough in the past months has been the implementation of a rotation mode for Starbugs. The fundamental design of the 'discrete stepping Starbug' had only ever accommodated translation in ±*x* and ±*y* directions, but the invention of a new sequence of actuator movements has now added angular positioning to the mini-robots' repertoire. Preliminary testing has shown that this technique is reliable and repeatable, with angular speeds in excess of 55 deg/s under nominal operation and a minimum angular step size of approximately 4 arcmins.

The freedom to arbitrarily rotate individual Starbugs brings new possibilities for the positioning system: optimised routing and collision avoidance with lower reconfiguration times; precise angular placement of non-symmetric optical payloads (e.g. alignment of IFUs); and field rotation compensation and tracking during observations.

### Performance data

Baseline performance data for the current vacuum Starbug design has been obtained using a dedicated laboratory 'test rig', shown in **Figure 6**. The test rig has control electronics for up to 20 Starbug devices and a 400 mm diameter interchangeable glass field plate. A high resolution machine vision camera is mounted above the plate and provides positional feedback for control software running on a desktop PC. The entire system is mounted on a motor-driven axle to simulate the movement of a telescope from zenith to the horizon.

Starbugs require four waveforms to operate, which have a specific shape in order to produce the discrete stepping motion. A Bug's direction of movement is set by applying these waveforms to different electrodes on the piezo actuators. The step size and movement speed of a Starbug can be controlled by changing the amplitude and frequency of the waveforms.

**Figure 7** shows the relationship between waveform frequency and Starbug speed. There is a clear linear range up to 150 Hz. Frequencies up to 300 Hz, though more prone to perturbations, yield speeds approaching 5 mm/s and can be effective when covering large distances. Over time, a nominal waveform frequency of 100 Hz has been empirically established as a balance of speed, accuracy and low power consumption. **Figure 8** shows how very fine positioning can be achieved by lowering the Starbug's drive voltage and reducing its step size.

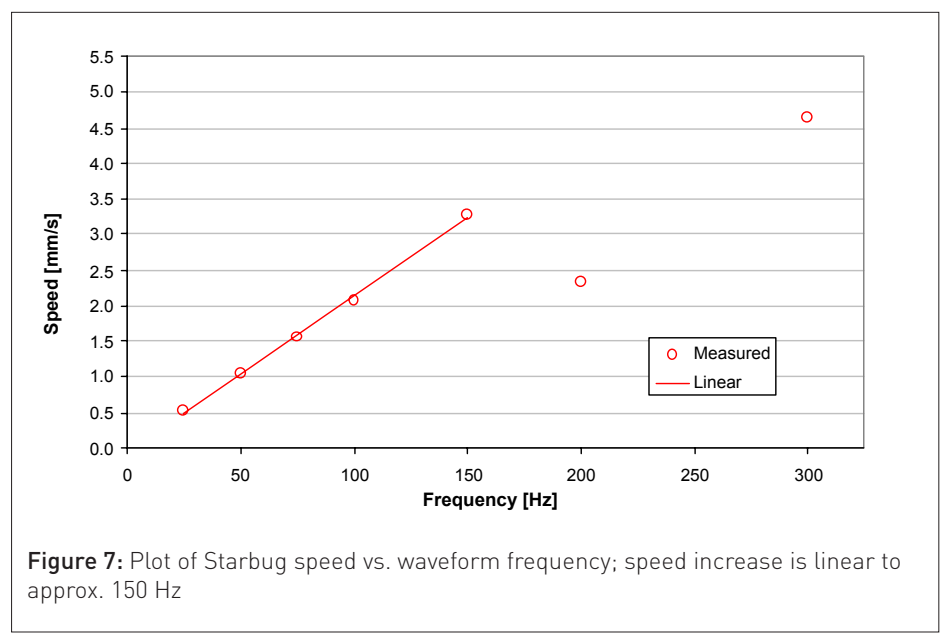

Figure 7: Plot of Starbug speed vs. waveform frequency; speed increase is linear to approx. 150 Hz

**Figure 9** shows that tilting the field plate through 90 degrees results in a small and approximately linear decrease in performance. These results are for a ~0.5 m long bundle of seven copper wires, four optical fibres and a silicone vacuum hose hanging from the Starbug.

### Minimum separation

The current generation of Starbugs for MANIFEST have a diameter of 8 mm, which, with a space cushion, gives a minimum separation of ~10 mm. This sizing produces a good holding force (due to the area under vacuum) and meets the target performance figures for the MANIFEST positioner. Previous prototype Starbugs as small as 6.3 mm in diameter have performed adequately; Starbugs smaller than this are a possibility, although at a cost of speed and payload size.

### Current work

At the time of writing we are close to completing the first parallel positioning test of 10 vacuum type Starbugs. A similar test was conducted with the old magnetic type Bugs in March 2011. We expect significant performance improvements with the new demonstration, especially at large field plate angles. This task has been undertaken by AAO summer student Alex Bennet. Alex is also assisting Michael Goodwin with earthquake simulation tests on the Starbug/vacuum system.

### Future work

Planned work for the near future will encompass further development of the MANIFEST Starbugs system and Starbug technology as a whole. Tasks will include:

- **Lifetime wear tests:** simulation of a Starbug's mechanical wear over the life of an instrument
- **Temperature effect tests:** analysis of ambient temperature effects on Starbug performance
- **Field plate optical coating tests:** measuring scratch/dig after long term use
- **New Starbug designs:** testing of larger Starbug variants for moving large fibre bundles / IFUs
- **Pick-off mirror integration:** investigation of Starbugs' suitability for pick-off mirror systems
- **Optimisation of electronics:** reduction of the control system's size, cost, power and heat dissipation

## Starbug vital statistics

Key performance figures for the current Starbug design for MANIFEST:

- Degrees of freedom: $x, y, \theta$
- Minimum x-y step size: <1 μm
- Minimum angular step size: 4.1 arcmin
- Maximum angular speed: >55 deg/s
- Maximum speed at zenith: >4.5 mm/s
- Maximum speed at 90 degree tilt: >4.0 mm/s
- Nominal speed at zenith: 2.1 mm/s
- Nominal speed at 90 degree tilt: 1.8 mm/s
- Minimum separation (pitch): ~10 mm
- Mass without wiring/hose: 3.5 g
- Mass with wiring/hose: 5.6 g

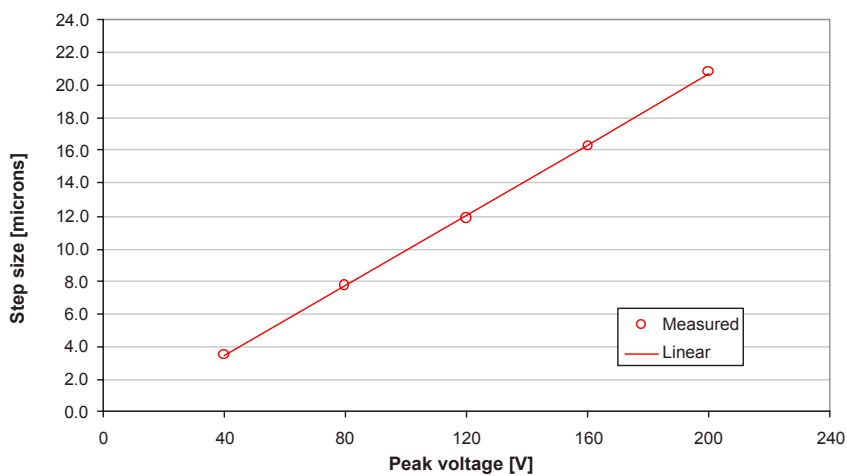

**Figure 8:** Plot of Starbug step size vs. waveform amplitude; results show an expected linear relationship

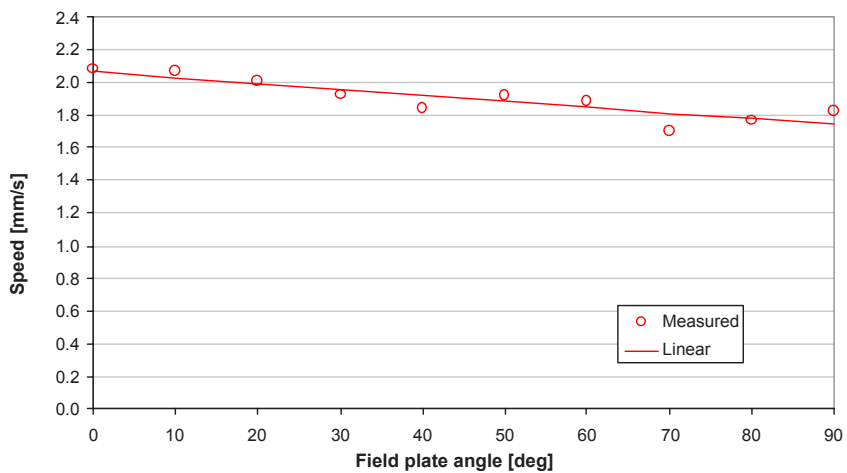

**Figure 9:** Plot of nominal Starbug speed vs. field plate angle for a ~0.5 m long fibre/wire bundle

# A message of progress from HERMES

Anthony Heng (AAO) and the HERMES Team

## Introduction

It has been a challenging and rewarding year for the HERMES project team. The team has been working tirelessly to produce a great instrument and to meet the project milestones. The VPH gratings are the most technically challenging component of HERMES, and only a couple of manufactures were close to having the ability to produce them. The optical team worked together with experts from the potential vendors to successfully overcome problems in the design of the gratings and to ensure they could be manufactured. Prototypes were procured and tested and the actual gratings have now been ordered. In other areas we experienced some of the problems that most projects face at some point—delays in the delivery of parts from vendors. This article will provide a snap shot of the HERMES project in terms of the design, procurement and the Assembly, Integration and Testing (AIT) documentation.

## Design

The detailed design of all significant components is now complete, with the more than 1000 detailed drawings required for manufacture of components produced. The image above shows the model of the fully assembled spectrograph. Only minor items such as assembly jigs, components for testing, etc. remain outstanding.

## Procurement

All significant items have now been ordered. We have already received the fold mirrors, three of the four detectors, the Hartmann shutters, the main shutters (Bonn shutters), all fibre cable components, most electronic components and circuit boards, and a vast range of smaller items. The first parts of the spectrograph frame have been received. Most of the major frame panels have been cut-out and have been sent for detailing.

Optical components other then the VPH gratings had been ordered a long time ago, and are currently being manufactured. There have been delays with some of these at our external vendor, but we have been able to modify our plan to work around most of this delay. The Infrared arm detector has now been selected and ordered.

Assembly of electronic circuit boards is well underway, with all CCD controller boards assembled and many instrument control boards complete.

One major problem over the next couple of months will be finding some space to store all of this stuff as it arrives! Currently we are using the Epping Instrument assembly lab, but we will need the space in there to actually assemble the thing—so a shipping container has been placed at the back of the Massey building to use as a store room. We look forward to having the spectrograph base frame completely constructed by the end of March.

Up at the AAT, the HERMES room main structure has been constructed in the 4th floor Coudé West Room. Installation of the 2nd tier floor structure, air-conditioning and ventilation systems, and power arrangements will be undertaken shortly. Construction of the fibre bundles is well underway, with AAOmega slitlets installed on the AAOmega part of the bundle.

## AIT – Assembly, Integration and Test

The Assembly, Integration and Testing documentation set describes how we will put the instrument together, align it, and test it is working correctly. The AIT documentation for HERMES project is an ongoing effort, with 70% of the documentation now complete and internally reviewed. The documentation will continue to evolve with input from final designs.

## Software

The Instrument Spectrograph Control Task, working with an instrument simulator, has been integrated into the 2dF control task. The 2dF control task can now run in a HERMES mode and talk to 4 CCDs rather then just the two of AAOmega.

The 2dFdr data reduction software is now reducing simulated HERMES images. Much remains to be done, but it is progressing well.

## Conclusion

2012 will be another exciting and challenging year for the HERMES project team, as we assemble, integrate and test HERMES at Epping, prior to it being sent to the AAT for its first commissioning run in March 2013.

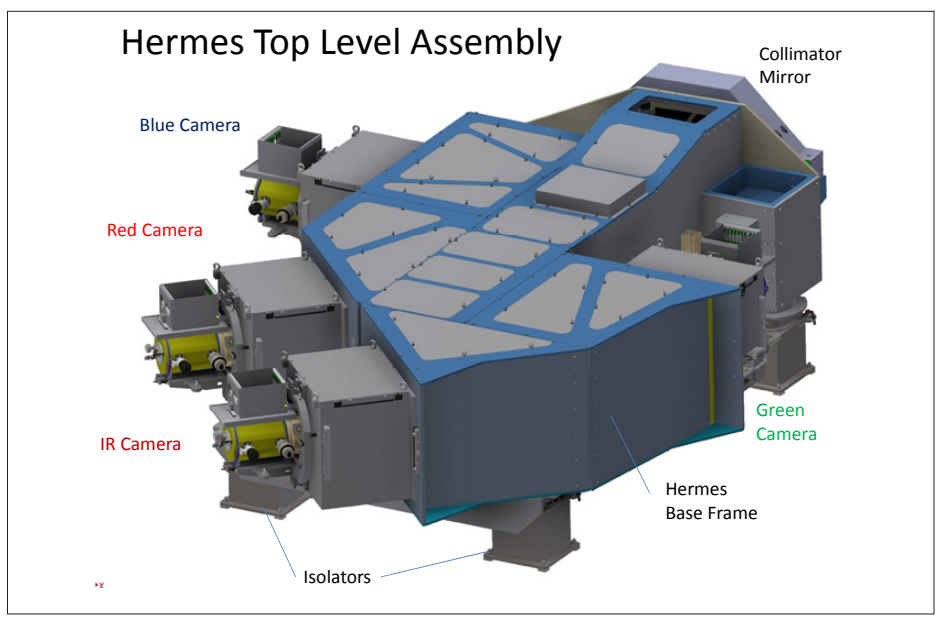

Hermes Top Level Assembly

- Blue Camera
- Red Camera
- IR Camera
- Isolators
- Collimator Mirror
- Green Camera
- Hermes Base Frame

# Imaging with the 2dF Focal Plane Imager

Ángel R. López-Sánchez (AAO/Macquarie)

*The 2dF FPI is a decent astronomical camera on a 4 metre telescope (see Dobbie et al., AAO Observer August 2011). Although it cannot compete with dedicated imaging instruments, it can easily do both target-of-opportunity science and poor weather science, as it is always available whenever 2dF is on the telescope. Below is an example of what can be done even in marginal weather. —Editor*

**The center of the interacting galaxies NGC 4038 and NGC 4039, "The Antennae", as it seen using the Focal Plane Imager (FPI) camera of the 2dF instrument on the AAT.**

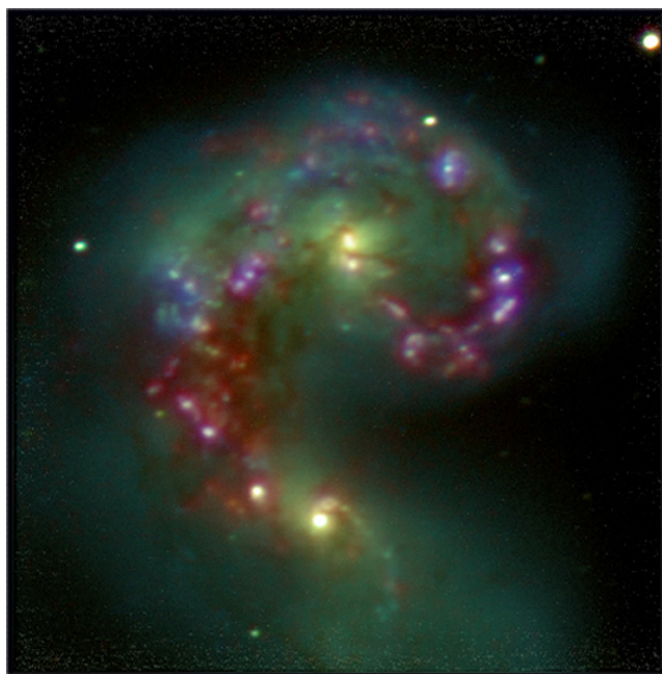
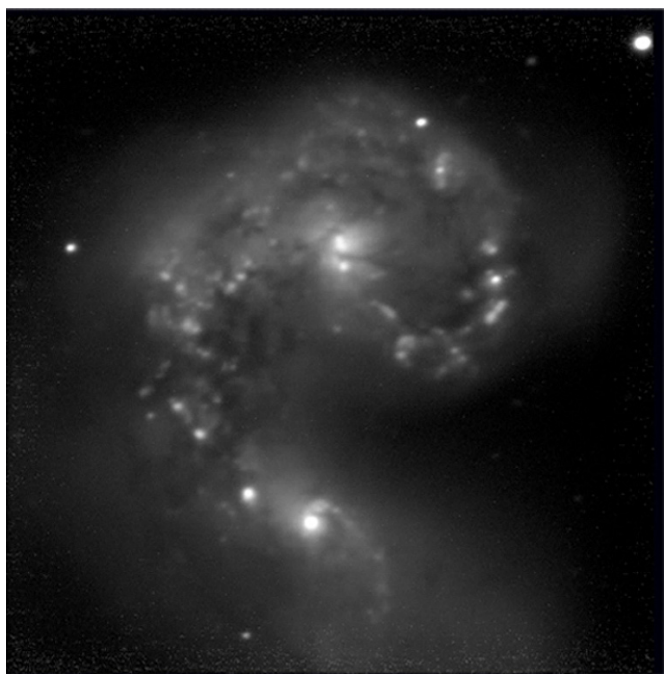

**(Left)** The colour image was obtained combining archive data of the interacting galaxies in several broad-band and narrow-band filters in UV and optical wavelengths. In particular, I used Far-UV data (dark blue) from the GALEX satellite, H-alpha data (red) from the 1.5m Palomar telescope (USA), and V (pale blue) and I (green) data obtained using the 2.5m telescope at Las Campanas Observatory (Chile). The luminosity of the colour image is given by the greyscale image obtained using the FPI camera of the 2dF instrument at the AAT.

**(Right)** Combined greyscale image. The raw images were obtained on the night of the 18th to 19th February 2012 (around 17h UT), while I was the support astronomer at the AAT. The weather at Siding Spring Observatory was very poor, and the target fields for the AAT service program that I had configured in both 2dF plates were completely blacked out by dark clouds. Although I took several images with exposure times between 1 and 5 min, only three exposures of 5 minutes each were used to produce this image. The alignment of the raw images was done using IRAF, but no bias or flatfield correction was applied.

# An Auspicious Launch

Russell Cannon (AAO) and David Malin (AAO)

The Australian Federal Government announced in 2009 that it would make up the shortfall in the AAO budget when the British formally withdrew. This was a real milestone for the Observatory (see messages from Matthew Colless in *Newsletters* **116**, August 2009 and **117**, February 2010). We realised that this event had to be appropriately celebrated and quickly settled on marking the occasion with a conference, to be held in Coonabarabran in June 2010 (article by Sarah Brough in *The AAO Observer* **119**, February 2011). On 1 July 2010 the original Anglo-Australian Observatory became the Australian Astronomical Observatory, timing that coincided neatly with the 35th anniversary of the start of scheduled observing on the AAT. Matthew Colless also suggested that we write a book as a continuation of "The Creation of the AAO" by Ben Gascoigne, Katrina Proust and Mac Robins (CUP 1990). Matthew looked hard at us as he said this, but we both immediately realised that writing such a history would be an enormous task and certainly not something we could complete on any reasonable timescale. As a compromise we agreed to edit the proceedings of the Coonabarabran meeting, but in a somewhat different style from a conventional scientific meeting. Our aim was to produce a book that would capture the stories of many of the people who had built, operated and used the AAT and the UK Schmidt Telescope (part of the AAO since 1988). We wanted to cover the main astronomical and technical highlights, but also to include less formal recollections of people and events, and we wanted to include many historical photographs. We also wanted an attractive and substantial book that we hoped would be mostly comprehensible to a wide readership, a resource that could provide a starting point for any future true historians.

The result was *Celebrating the AAO: Past, Present & Future*. Although the bulk of the writing was done by over 50 co-authors, compiling, editing and checking the final result still took well over a year. The then DIISR (Department of Industry, Innovation, Science and Research) undertook to publish the book, which was eventually finished in September 2011

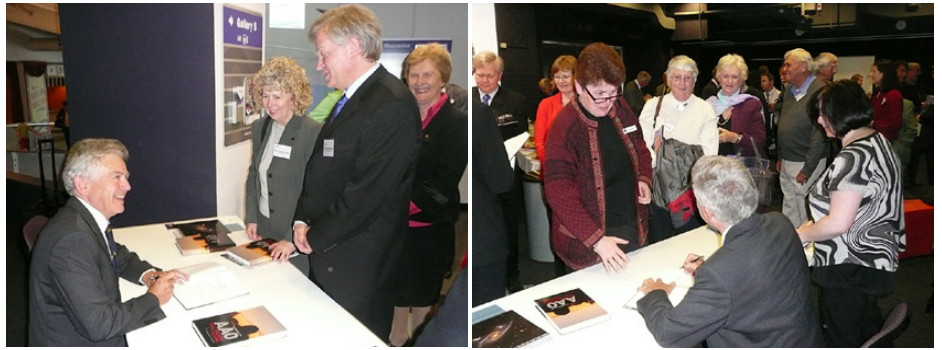

(left) Russell Cannon with Brian Schmidt and Elaine Sadler; **(right)** part of the queue.

(we particularly want to thank Russell Wilson for his advice and oversight of the project, including the key suggestion of using colour throughout). The final step was to organise a formal launch, for which Questacon kindly offered the use of its premises in Canberra. The event was set for 13 October, which turned out to be an auspicious day: a few days earlier we had learned that Brian Schmidt was a winner of the 2011 Nobel Prize for Physics, for his part in the discovery that the expansion of the Universe was accelerating. The Academy of Science decided to celebrate this recognition with an early morning champagne breakfast in the Shine Dome. As a result, some of us were invited to that ceremony, but more significantly a substantial group of Academicians and eminent Australians accompanied Brian to our book launch, which he attended as an AAO Advisory Committee Member.

The launch, ably organised by Cathy Parisi, was attended by 60 guests, including representatives from many Australian Government, scientific and educational organisations, as well as about 20 of the contributors to the book. AAO media star Fred Watson was MC and Russell Cannon spoke first, on behalf of the editors. Sadly, David Malin was overseas but had prepared a slide show, which ran in the background. Stuart Wyithe from Melbourne University, winner of the 2011 Malcolm McIntosh Prize for Physical Scientist of the Year, formally launched the book, while Andrew Hopkins represented the AAO (Matthew was also overseas) and Brian Schmidt represented the AAO Advisory Committee. The recent time-lapse film of the AAT dome by Angel Lopez-Sanchez (see article this issue) was shown at the end of the proceedings, followed by a book-signing by Russell while guests enjoyed refreshments provided by Questacon.

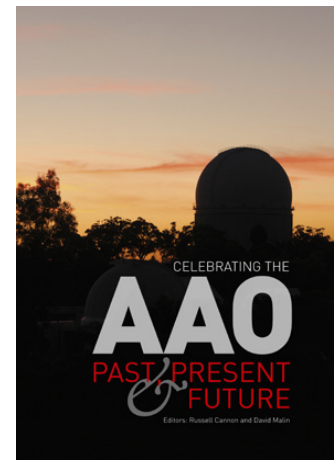

We hope that this book will fulfil at least some of our ambitions, and that readers will enjoy the mixture of factual accounts, reminiscences and sometimes slightly dubious anecdotes, as well as the abundant illustrations. In some ways the combination of many separate stories may give a truer picture of the AAO than any single coherent historical account. Copies of the book can be ordered from the AAO, using the order form at: *http://www.aao.gov.au/press/booklaunch2011/*

# AusGO CORNER

Stuart Ryder (Australian Gemini Office, AAO)

### A New Era for AusGO

Since January 2008 the Australian Gemini Office (AusGO) has been operated by the AAO under a contract with AAL[1] from funds provided by the National Collaborative Research Infrastructure Strategy (NCRIS). This funding covers partial salaries for the Australian Gemini Scientist and two Deputy Gemini Scientists; the cost of attendance at Gemini meetings for AusGO staff and Australian members of the Gemini Board, the Gemini Science Committee (now the Science and Technology Advisory Committee), and the International Time Allocation Committee; and discretionary activities like the recent Observational Techniques workshop and annual school astronomy contests.

With the conclusion of the NCRIS program in 2011, funding for AusGO will now be covered by the AAO's own operating budget, while meeting travel costs for non-AusGO staff will be covered by AAL. While most users should notice no significant difference in the quality of support they receive from AusGO, the change in funding channels does offer up an opportunity to refocus our efforts and reallocate resources if warranted. As outlined elsewhere in this issue, as part of the AAO Forward Look process, AusGO will soon be carrying out an extensive consultation process with its users and stakeholders to look at how we can best meet the needs of the Gemini and Magellan user community in the post-2012 period. We very much welcome your input and ideas as we seek to build an even better AusGO within the AAO.

### Proposal Statistics

For Semester 2012A the Australian Time Allocation Committee received a total of 32 Gemini proposals, of which 21 were for time on Gemini North, 1 for exchange time on Subaru, and 10 were for time on Gemini South. The oversubscription for Gemini North went up from 2.3 in 2011B to 3.2, while demand for Gemini South was stable at 1.3. The amount of science time available on Gemini South in 2012A was significantly reduced due to commissioning of GeMS and FLAMINGOS-2. Magellan time in 2012A was oversubscribed by a factor of 4.0, with 13 proposals.

AusGO hosted the 2012A meeting of the Gemini International Time Allocation Committee at the AAO on 18–19 Nov 2011.

In 2011A, all but one of the seven Band 1 programs were completed or had insufficient Target of Opportunity triggers; 4 of 5 Band 2 programs were completed; and 4 of 7 Band 3 programs were completed. Almost 90% of Australia's time went into completed programs.

### Magellan and Gemini travel funding

The ANSTO[2]-administered Access to Major Research Facilities Program (AMRFP), which has supported travel to overseas observatories for a number of years, finished in June 2011. DIISR gave approval to use some of the uncommitted international travel funds from the final 6 months of AusGO's NCRIS grant to cover Magellan observer travel in Semester 2011B. From Semester 2012A onwards, the AAO has allocated funding from its own budget to AusGO to ensure that users who are fortunate to be awarded classical time on Magellan, Gemini, or Gemini exchange time on Subaru or Keck are actually able to take up that opportunity. Reimbursement policies can be found at http://ausgo.aao.gov.au/magellan.html#travel. Qantas has announced that it will soon replace its Sydney – Buenos Aires service with a direct Sydney – Santiago service, which should add a healthy degree of competition on this relatively expensive route.

### Gemini High-resolution Optical Spectrograph

An AAO-led team is one of three selected by Gemini to carry out 6 month Conceptual Design studies for the Gemini High-resolution Optical Spectrograph (GHOS), a proposed next-generation instrument for Gemini. The other teams are led by the University of Colorado at Boulder, and the Herzberg Institute of Astrophysics. If a decision is made to proceed to the build phase, GHOS will have a nominal resolution of R~40,000, provide simultaneous wavelength coverage from 370 to 1000nm (with a goal of covering 300 to 1100nm), and provide stability of 1/3 pixel/hour (with a goal of 1/10 pixel/hour). Delivery of GHOS to Gemini South is targeted for 2015.

### Instrument Commissioning Update

The second week of December 2011 saw the recommencement of commissioning for FLAMINGOS-2 with its replacement science grade detector. Evaluation of its performance is still ongoing but the delivered image quality is excellent. A call for System Verification (SV) observations with FLAMINGOS-2 has now been issued. The following week it was the turn of GeMS, the Gemini MCAO System which comprises the Canopus optical bench, the 5 laser guide star constellation, and the RSAA-built Gemini South Adaptive Optics Imager (GSAOI). Characterising each system then integrating them together is a long, challenging process but already images have been obtained with 50 milli-arcsec FWHM in *H*, and Strehl ratios across the full field of 40% (see http://www.gemini.edu/node/11718). Meanwhile, despite atrocious weather on Mauna Kea keeping Gemini North closed for extended periods, the new E2V deep-depletion CCDs in GMOS-N have been declared ready for science use in "6 amp" mode (i.e. reading out half of each CCD through separate readout registers). Planning is now underway to install

---

1 Astronomy Australia Limited.

2 Australian Nuclear Science and Technology Organisation

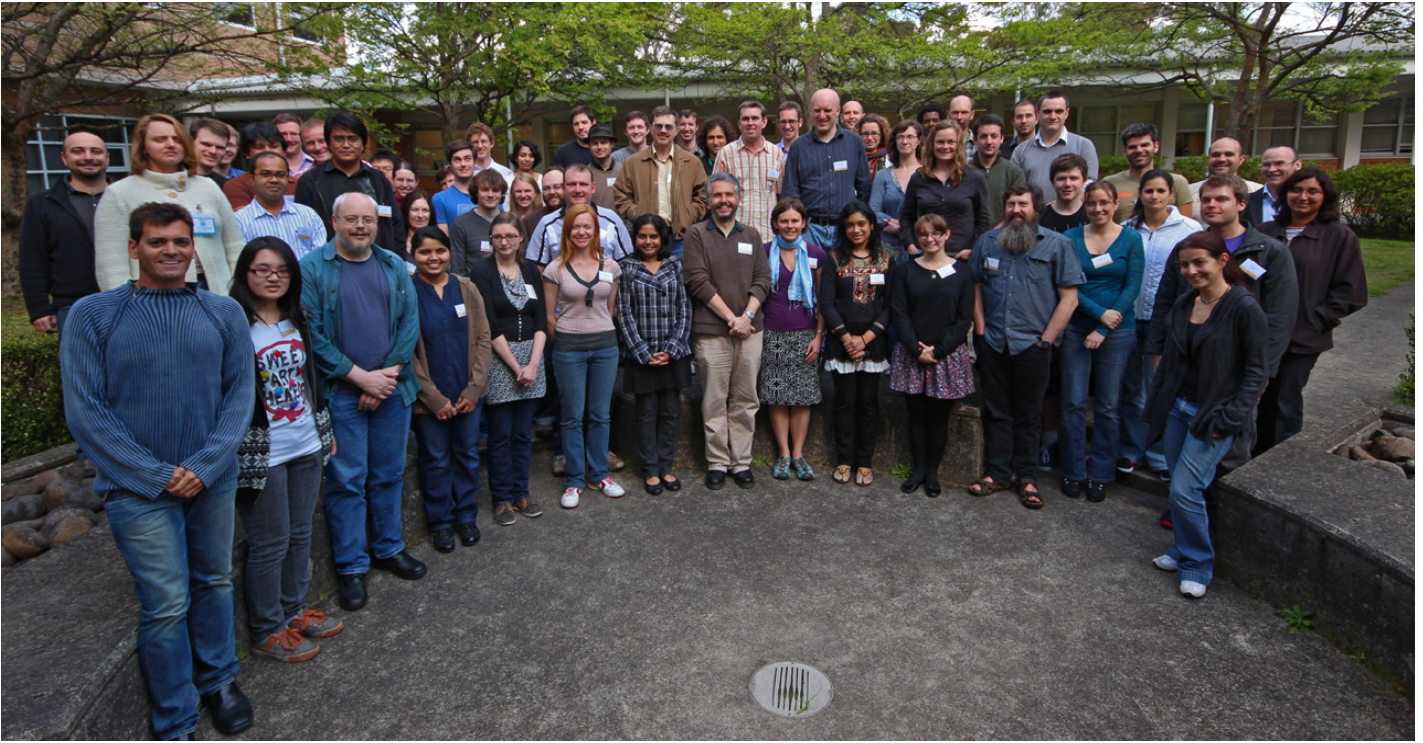

Figure 1: Participants at the 2011 AusGO/AAO Observational Techniques Workshop. Image credit: Angel Lopez-Sanchez

Hamamatsu CCDs with their superb quantum efficiency right out to 1 micron in both GMOS-North and GMOS-South by mid-2013. Thus the coming year will see a significant boost in instrumentation capabilities at Gemini, which we expect to translate into increased demand from the community.

## Observational Techniques workshop

AusGO, with assistance from other AAO staff members and the Gemini Observatory, organised a workshop on optical and infrared observational techniques, focusing on the facilities offered on the AAT, Gemini and Magellan telescopes. The workshop was held in the CASS[3] lecture theatre from 30 August to 2 September, and was attended by over 50 students and postdocs and some 20 speakers (see Figure 1). Emma Hogan travelled from Gemini South to provide assistance with installation and use of Gemini IRAF, while James Radomski gave a presentation on mid-IR astronomy by video from Gemini South. Talks also covered aspects such as astro-informatics, large surveys, and media releases. All presentations are available from the workshop web site (http://www.aao.gov.au/AUSGO-AAO_Workshop/). Feedback from participants has been very positive, and it is planned to hold such workshops every second year from now on.

3   CSIRO Astronomy and Space Science

## AGUSS

The Australian Gemini Undergraduate Summer Studentship (AGUSS) program is sponsored by AAL from its Overseas Optical Reserve. It offers talented undergraduate students enrolled at Australian universities the opportunity to spend 10 weeks over summer working at the Gemini South observatory in La Serena, Chile, on a research project with Gemini staff. They also assist with queue observations at Gemini South itself, and visit the Magellan telescopes at Las Campanas Observatory. The two AGUSS recipients for 2011/12 are Joe Callingham and Ayna Musaeva, both from Sydney University (Figure 2). Joe is working with Benoit Neichel and Claudia Winge, analyzing data to characterise the sodium layer above Cerro Pachon, while Ayna is working with Tom Hayward and Fredrik Rantakyro, collecting and analysing data regarding the image quality on Cerro Pachon. They will be presenting their results to the other AAO and CASS summer students in Sydney by video in early-February shortly before their return to Australia.

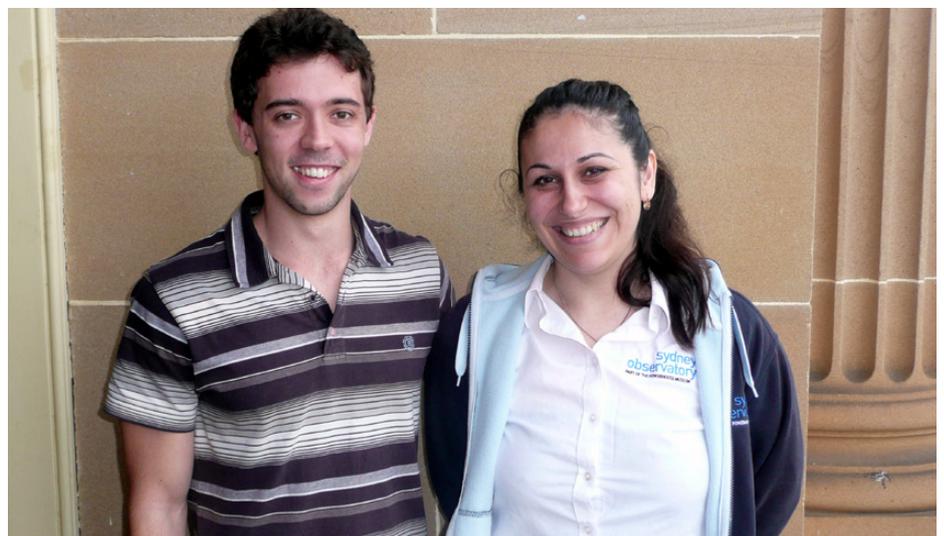

Figure 2: AGUSS recipients for 2011/12, Joe Callingham and Ayna Musaeva. Image credit: Stuart Ryder

# Gemini School Astronomy Image Contest Winner 2011

Chris Onken (ANU)

Benjamin Reynolds, a Year 10 student from Sutherland Shire Christian School, won the 2011 Gemini School Astronomy Contest. He proposed that Gemini South image the nearby, barred spiral galaxy, NGC 7552. The 2011 contest, the third such competition run by the Australian Gemini Office (AusGO), awarded Ben with a framed copy of this new image from Gemini's GMOS camera at a ceremony held at Ben's school in Sutherland, NSW.

The runners-up in the contest were Ryan Soares from Trinity College in East Perth, WA, and a group from St. Margaret's Anglican Girls School, Ascot, QLD, comprised of Eugenie Puskarz-Thomas, Rachel Augustyn, Phoebe Duncombe, Brooke Henzel, Matilda Williams, and Louise Graham. Students at all three of the schools were given the chance to learn more about Gemini through a videoconference with staff members at the Observatory headquarters in Hilo, Hawaii.

Separate observations with Gemini's mid-infrared instrument T-ReCS probed the central regions of NGC 7552. The images revealed a ring of dust heated by a recent episode of star-formation and gave a more complete picture of the possible fate of the gas and dust seen in the GMOS image.

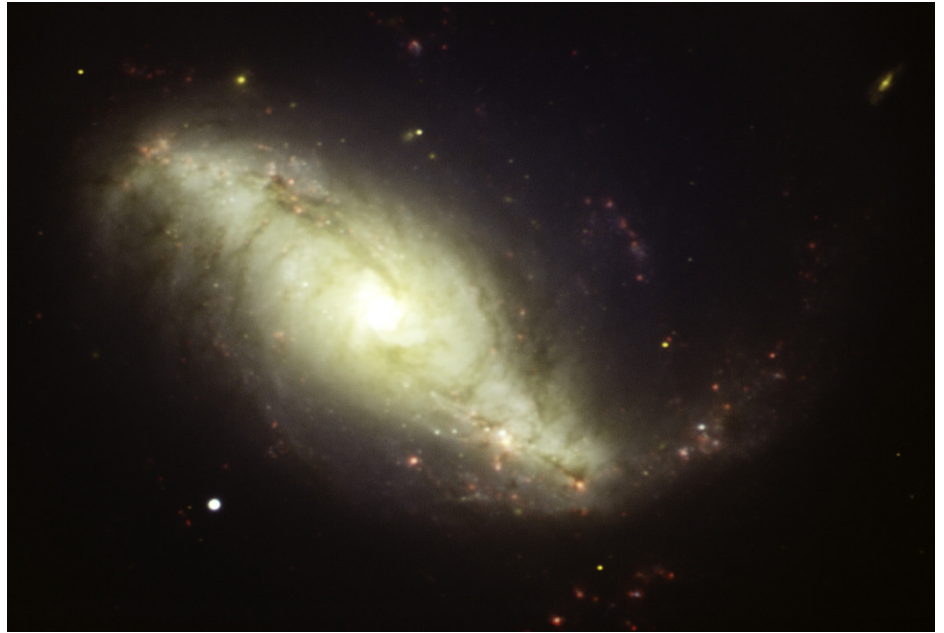

The winning image in the 2011 Gemini School Astronomy Contest featured the local spiral galaxy NGC 7552. The GMOS instrument on Gemini South captured the image for Sydney-area student, Benjamin Reynolds, whose contest-winning entry suggested the target. The resulting picture is a colour-composite created from H-alpha (red), and SDSS g, r, and i filters (blue, green, and yellow, respectively). Image credit: Benjamin Reynolds (Sutherland Shire Christian School), Travis Rector (University of Alaska Anchorage), and the Australian Gemini Office

AusGO is excited to be conducting a new contest for 2012. Details are available from the AusGO website: http://ausgo.aao.gov.au/contest/

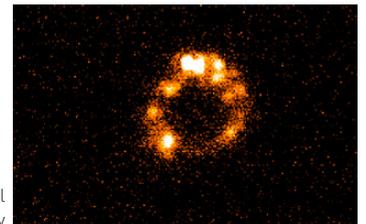

T-ReCS mid-infrared image of the central regions of NGC7552 Credit: Gemini Observatory

Credit: Gemini Observatory

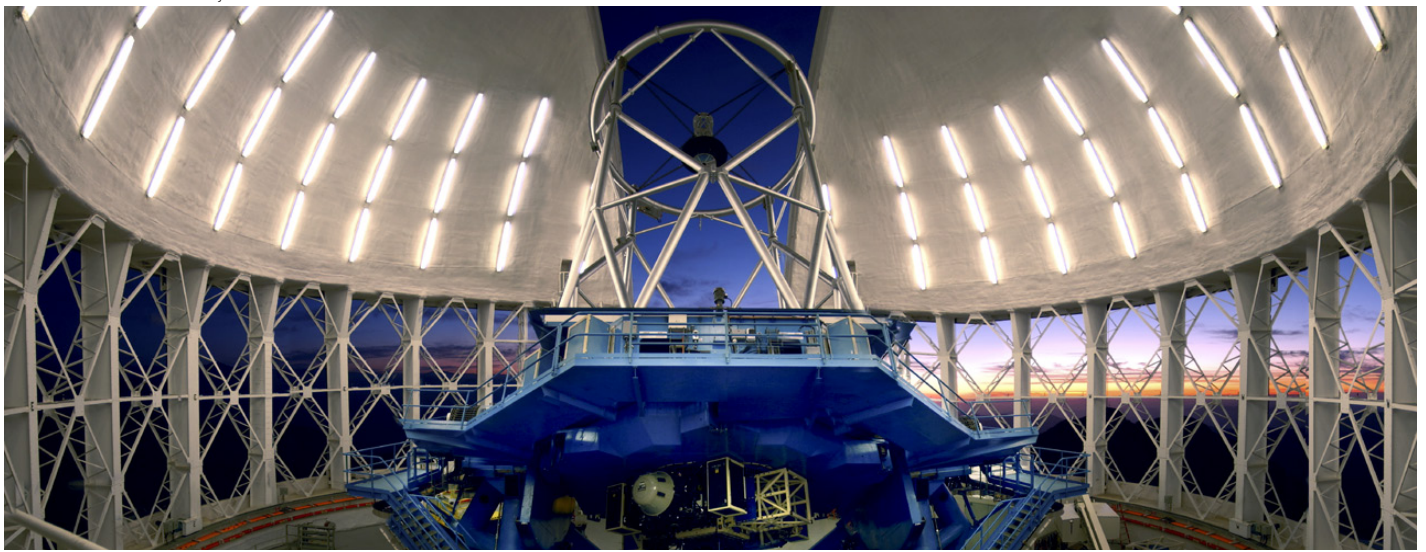

# The Sky over the Anglo-Australian Telescope

Dr. Ángel R. López-Sánchez (AAO/Macquarie)

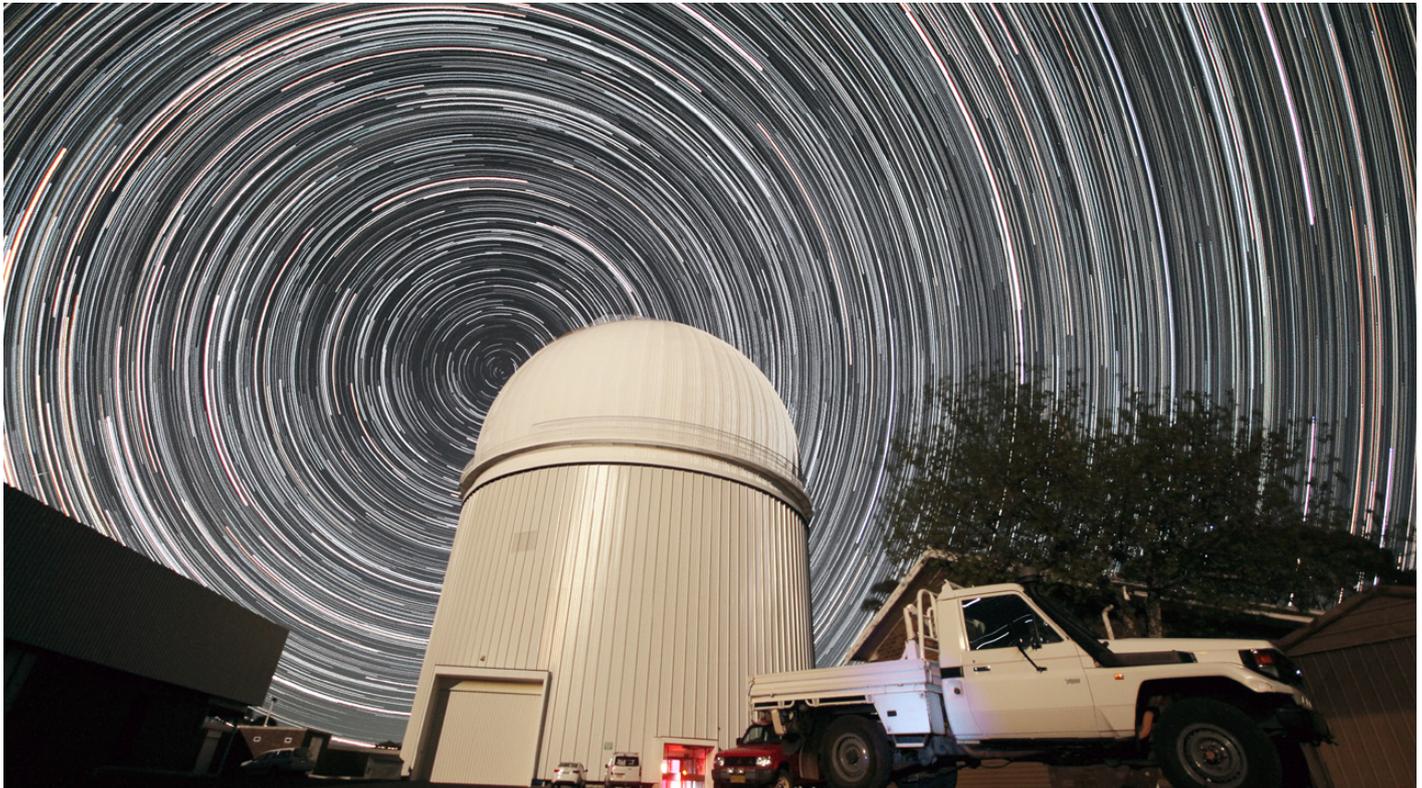

Circumpolar star traces (2.7 hours) over the Anglo-Australian Telescope on 20 Sep 2011. Credit: Dr. Ángel R. López-Sánchez (AAO/Macquarie Uni.)

A dark winter night, with the Milky Way crossing the firmament while its center in located near the zenith, is one of the most astonishing views we can enjoy. This view is only available from the Southern Hemisphere and it is quite inspiring. In particular, the Milky Way shines over the Siding Spring Observatory, near Coonabarabran, New South Wales, where the famous AAT is located. With the idea of sharing the beauty of the night sky to everybody, I decided to start taking "time-lapse photography" while I was working as support astronomer at the AAT. This technique consists on taking many images and then combining them to get a very high-resolution movie. The best shots I obtained so far have been included in the video "*The Sky over the Anglo-Australian Telescope*", that you can access through this internet address:

http://www.aao.gov.au/press/timelapse/

This video, which lasts for 2.7 minutes, is the results of combining around 3800 different frames obtained using a CANON EOS 600D between June and September 2011. Except for those frames used for the sunset in the first scene, all frames have a 30 seconds exposure time, with a ISO speed of 1600. As the videos were created at 24 fps (frames per second), each second in the movie corresponds to 12 minutes in real time. I used several lens to take the images (standard 50 mm, 50mm x 0.65 focal reducer and a 10 mm wide-angle lens). The aperture chosen was 5.6 (for the 50 mm lens) or 4.5 (10 mm wide-angle lens). Processing each sequence of the movie took five to six hours of computer time, and usually I had to repeat this at least once for each sequence, to improve the quality. The soundtrack I chose is an extract of the music "*Echoes from the past*", by the French composer Dj Fab, which gives energy to the time-lapse.

As my main job while I'm at the AAT is giving support to the astronomers who are observing in this telescope, I always set the camera up at the beginning of the night, let it run, and check on its progress occasionally. Sometimes this was not easy: wind knocked the camera over on a couple of times, often the battery ran out, and even once I had an encounter with some intransigent kangaroos. However, finally I got this material, which not only shows the magnificent Milky Way rising and setting above the dome of the AAT, but also stars circling the South Celestial Pole, the Magellanic Clouds over the AAT, satellites and airplanes crossing the sky, the Moon rising and setting, and the most famous constellations of Orion, Carina and the Southern Cross. I hope you enjoy the result.

**Dr. Ángel R. López-Sánchez**
Australian Astronomical Observatory and Macquarie University Research Fellow
P.O. Box 296, Epping, NSW 1710, Australia
PH: +61 2 9372 4849,
FAX: +61 2 9372 4880
http://www.aao.gov.au/local/www/alopez/

# New Faces at the AAO

Andy Green (AAO)

## New Faces

There are several new faces around the offices in Epping. Five new astronomers have joined the AAO, plus the new Head of Instrumentation, Andy Sheinis.

**Dr Matt Owers** grew up in Thirroul, which is a suburb of Wollongong in NSW. After a couple of false starts in Engineering, he completed an undergraduate degree in Physics at the University of Wollongong before moving the University of New South Wales to undertake a PhD with Warrick Couch. Upon completion of his PhD in 2008, Matt followed Warrick to Swinburne for his first postdoc. To date, his research has focused on combined X-ray and optical studies of merging cluster of galaxies, with some excursions into the environments of starburst and poststarburst galaxies. Matt started at the AAO in September 2011 as a Super Science Fellow and will be working with the GAMA collaboration with the broad aim of understanding the effects of environment on the star forming properties of galaxies. Away from astronomy, the majority of his time is spent playing, watching or talking about cricket.

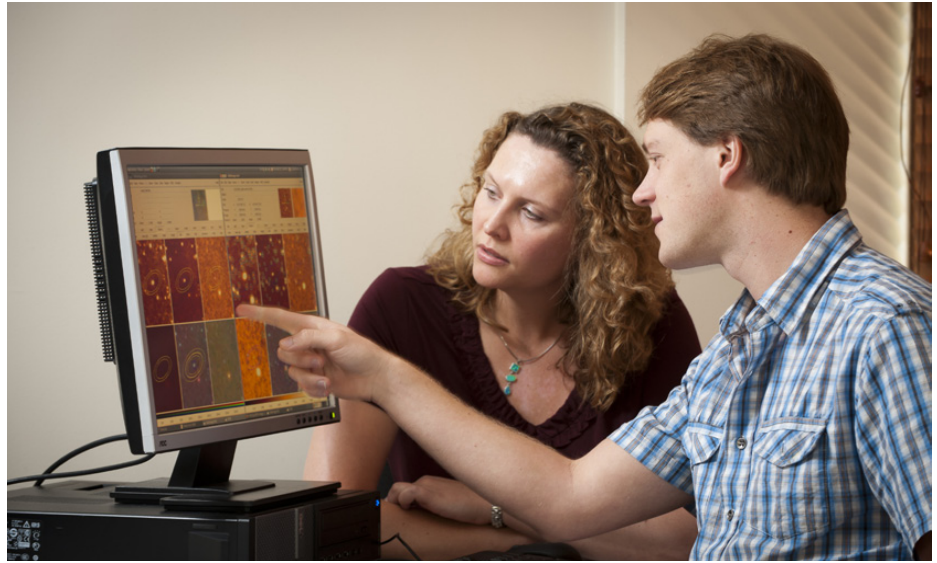

Michelle Cluver and student Kieran Leschinski

**Dr Michelle Cluver** joins the AAO as a Super Science Fellow. She comes from the Spitzer Science Center at Caltech. Michelle obtained her PhD from the University of Cape Town in 2009 and has experience with multiwavelength imaging and spectroscopy, from radio through far- and mid-IR to optical and UV measurements. Her research interests include evolution within galaxy groups, and the fueling of star formation as traced by HI and dust emission. Her current focus is on galaxy evolution analyses using the GAMA and WISE survey data.

**Dr Sarah Martell** is originally from the US: she grew up in the Northwest, and completed her PhD at UC Santa Cruz. For the last three years, Sarah has been working with Eva Grebel at the Astronomisches Rechen-Institut in Heidelberg, Germany, before becoming an AAO Research Fellow in October. She is an observational astronomer, with interests in stellar abundances, galactic archaeology and globular clusters. At the AAO, she will be supporting CYCLOPS, UCLES, and the UHRF, and is also taking over the role of AAT Scheduler from Paul Dobbie. Sarah lives in Ryde with her husband and their five-year-old son. Sarah likes to knit, read, and spend time by the ocean.

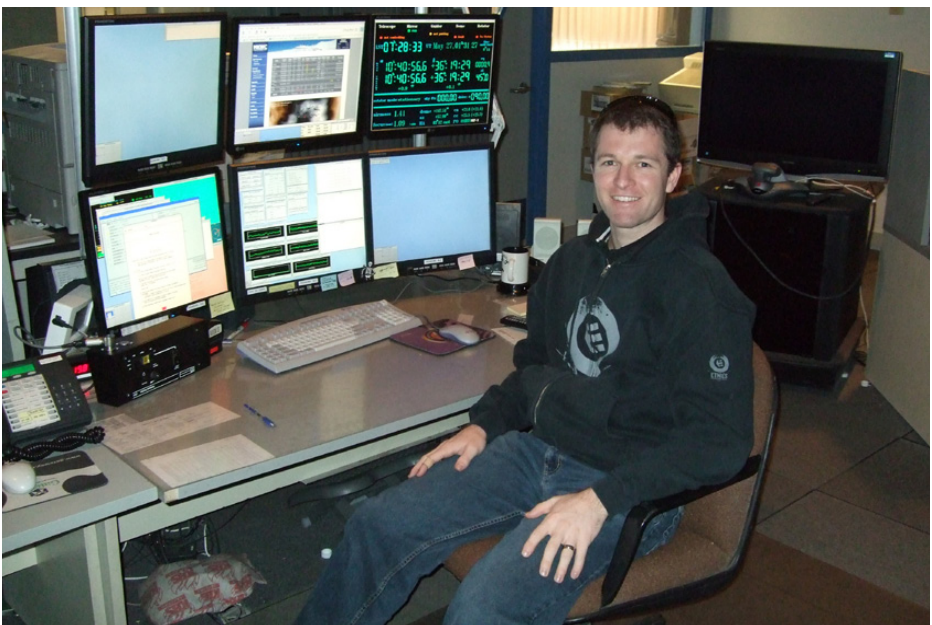

Dr Matt Owens

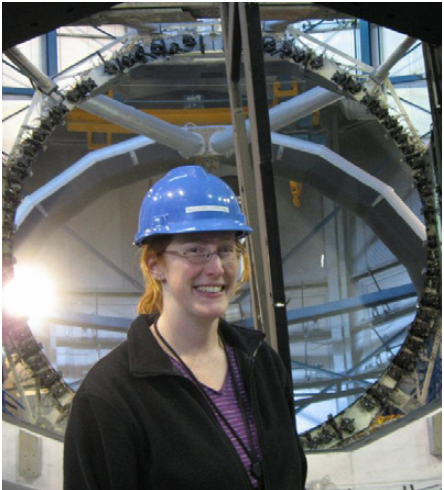

Dr Sarah Martell

**Dr Francesco di Mille** was born in Gaeta, Italy: a small town on the Tyrrhenian Sea south of Rome. At the University of Padua he received his first degree in astronomy (Laurea) in 2002, and his PhD in 2007. Before completing the PhD, Francesco joined the Asiago Astrophysical Observatory, first as resident astronomer and then directed the refurbishment of the 1.2m telescope. In 2009 he moved to Chile to join the Las Campanas Observatory as an Australian Magellan Fellow, the final year of which he is now completing at the AAO and Sydney University. Francesco's interests include the extended emission line regions around Active Galactic Nuclei, the role of galaxy interaction on nuclear activity, extragalactic symbiotic stars and classical novae. Beyond work, he enjoys time with wife Mariantonietta and child Antonio. Francesco loves everything related to the sea: swimming, snorkeling or just walking on the beach.

**Dr Andy Green** is originally from Colorado, completed his undergraduate degree at Caltech and has just finished his PhD with Karl Glazebrook at Swinburne University in Melbourne. Andy has moved to Sydney as an AAO Research Fellow, and will help support the AAOmega instrument suite. He is also taking over the AAO newsletter from Sarah Brough. Andy's astronomy interests are primarily in the formation and evolution of star forming galaxies studied via integral field spectroscopy, as well as public outreach. Outdoors, he enjoys climbing, mountain biking, tramping/bushwalking, cross-country skiing, and is keen to start sea kayaking around Sydney.

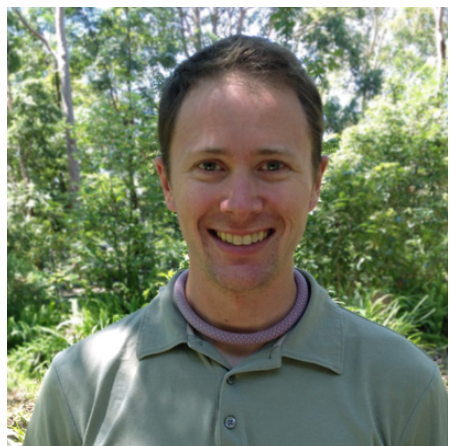

Dr Andy Green

**Professor Andy Sheinis**, the new head of Instrumentation, comes to the AAO with over 25 years experience in instrumentation in academia and industry, including developing instruments for the SALT, Keck, Lick and the AEOS telescopes. In addition, Andy is an active observer and studies distant galaxies and their relationship to

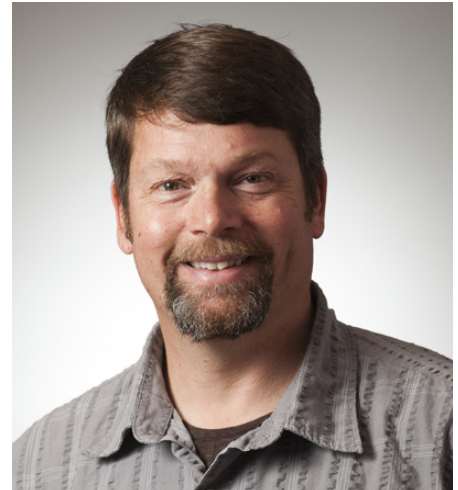

Professor Andy Sheinis

the super-massive black holes lurking in their cores. Before coming to the AAO, he was on the Astronomy faculty at the University of Wisconsin. Prior to that he was a National Science Foundation Postdoctoral Fellow at the Center for Adaptive Optics in Santa Cruz CA. He earned a PhD and M.S.. in Astrophysics from Lick Observatory at the University of California at Santa Cruz, where he was the Bachmann Instrumentation Fellow. He holds an M.S. in Optical Physics from Worcester Polytechnic in Worcester MA (USA) and a B.S. (with Honors) in Physics from the University of Massachusetts at Amherst (USA). He holds 3 US patents with 4 more pending in optical/IR technology, and has authored or co-authored over 40 scientific/technical articles, and teaches a regular invited course in spectrograph development at the SPIE. When he is not building instruments and doing astronomy, Andy enjoys mountain biking, camping, hiking (a.k.a. bushwalking), surfing, sailing, playing the guitar and ukulele badly, painting and sculpture. He used to enjoy ice skating and skiing, but it is unclear how much of that he'll be doing in Sydney. AAO

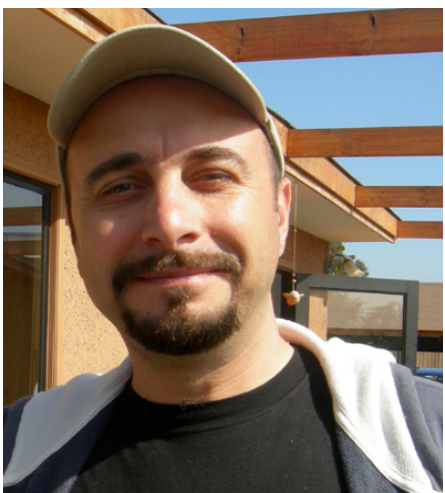

Dr Francesco di Mille

# AAO Distinguished Visitors

## 2012
### AAO Distinguished Visitors

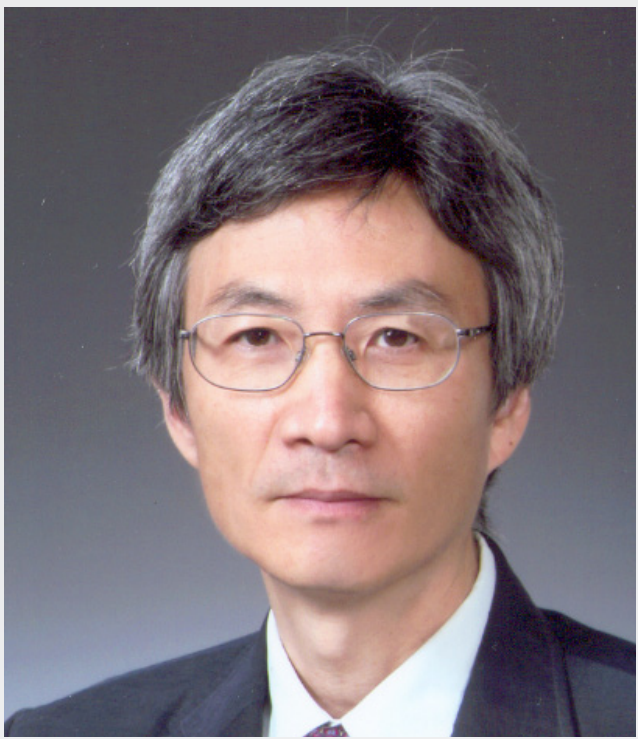

**Professor Bon-Chul Koo**, (Seoul National University), visiting March-May 2012.
Prof Koo will be working with AAO staff on supernova remnants and luminous blue variable stars in the Milky Way and the Large Magellanic Cloud, using data from the IRIS2 instrument on the AAT. Prof Koo is hosted by Andrew Hopkins.

**Dr Francesca Primas**, (ESO), visiting June-July 2012.
Dr Primas will be working with AAO staff on understanding stellar chemical abundances and their impact on planet formation within the Milky Way, using high resolution optical spectroscopy. Dr Primas is hosted by Gayandhi De Silva.

Since 2010, the AAO provides an annual Distinguished Visitor scheme. The aim of this scheme is to strengthen and enhance the AAO's visibility both locally and internationally, and to provide opportunities for AAO staff to benefit from longer term collaborative visits by distinguished international colleagues. The AAO is pleased to host three Distinguished Visitors during the first half of 2012.

Each of our Distinguished Visitors will be giving colloquia both at the AAO and at other institutions during their visit, to highlight their work and collaborations with the AAO. They will also engage with the general public through public lectures or other outreach activities. We encourage members of the astronomical community to visit the AAO and take advantage of these visits by our distinguished colleagues to build new collaborations, to reinforce or rekindle existing ones, and to maximise the outputs from their time with us at the AAO.

**The next deadline for Distinguished Visitor applications will be 31 March 2012, for visits during July 2012 – June 2013. Details of the AAO Distinguished Visitor Scheme can be found on the AAO web pages at: http://www.aao.gov.au/distinguished_visitors/**

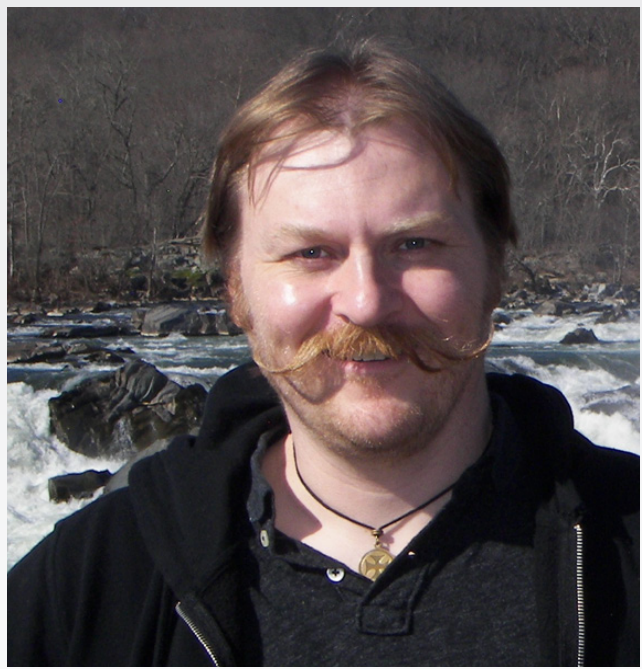

**Associate Professor Christopher Stockdale**, (Marquette University), visiting March-June 2012.
A. Prof. Stockdale will be working with AAO staff on monitoring and understanding the origins of supernovae using observations from Gemini and ATCA. A Prof Stockdale is hosted by Stuart Ryder.

# Epping News
Andy Green (AAO)

## Staff news

**Dr Paul Dobbie** has left the AAO, after completing his term of appointment as AAO Research Astronomer. Paul made substantial contributions to the AAO through his roles as Newsletter editor and Scheduler as well as instrument scientist for IRIS2. He also provided high quality support for AAOmega, and was regularly acknowledged explicitly by observers in their feedback. Paul has taken up a Research Fellow position at the University of Tasmania, from December 2011.

**Dr Max Spolaor** moved to the US in August 2011, to take up a position at UCLA studying atmospheric chemistry. As part of a NASA-funded project, he will be responsible for an instrument on board a NASA Global Hawk unmanned high-altitude aircraft. In his short time at the AAO, Max played an important role as the AAOmega instrument scientist, taking over from Rob Sharp, and contributed in advancing the project for the AAOmega upgrades.

The AAO wishes both Paul and Max the very best in their new positions.

**Dr Dan Zucker** has been awarded an ARC Future Fellowship, hosted at Macquarie University. We are pleased that Dan will continue as an Honorary Associate with the AAO during his Fellowship. Macquarie is presently advertising for a 4 year joint appointment between Macquarie and AAO to replace the AAO/Macquarie Lecturer position that Dan leaves vacant. We look forward to continuing a close interaction with Dan during the duration of his Fellowship.

## Student Fellowships

The AAO continues its program of 10-12 week undergraduate fellowships. These are supervised by an AAO staff member, and are highly competitive, with an oversubscription rate of 5–20. More information is available from the AAO website: http://www.aao.gov.au/AAO/students/aaosf.html

Three Student Fellows began projects in December 2011:

**Kieran Leschinski** (Monash, Melbourne) is working with Michelle Cluver on the project "A Pilot Study of Resolved WISE Sources in GAMA." This project is to conduct photometric measurements of extended sources in the WISE (Wide-Field Infrared Explorer) satellite imaging, combined with the other multi-wavelength photometry, to explore the full ultraviolet-to-far-infrared spectral energy distribution of these galaxies.

**Adam Stevens** (Canterbury, New Zealand) is working with Stuart Ryder on the project "Stellar and Gaseous Abundance Gradients in Galaxy Disks." This project uses Gemini spectroscopy to explore the abundance gradients of iron and magnesium, to compare against that of oxygen, for a sample of spiral galaxies, in order to reconstruct their star formation and chemical enrichment histories.

**Janette Suherli** (Bandung, Indonesia) is working with Chris Lidman and Sarah Brough on the project "Is there evolution in the BCG population in high redshift clusters?" This project explores the properties of the brightest galaxies in high-redshift clusters identified with Spitzer, compared against a low-redshift sample, in order to test constraints from galaxy formation models.

Additionally, we have two Instrument Science Studnets: **Alex Bennet** (UWA) is working with Michael Goodwin and Jamie Gilbert on Starbugs (see article this issue). He has helped characterise the newly integrated rotation mode, assisted in the recent parallel positioning demonstration, and conducted earthquake testing. **Jendi Kepple** (UNSW) is working with Simon Ellis on modelling ring resonators for astronomy.

## Instrument News

Sarah Brough, Angel Lopez-Sanchez and Andy Green have been working hard on the **AAOmega** documentation. Both the AAOmega web pages and much of the written documentation have been extensively reviewed, edited and updated, and they expect to release these improvements soon. Also, Mike Birchall is in the late stages of testing a new release of the 2dfdr data reduction tool, which should be both more robust and user friendly. The 2dF robot has received some work from Rob Patterson and Darren Stafford, resulting in a significant drop in configure time: 4–6 seconds per fibre, and generally 40–45 minutes for a complete field (a 25–35% improvement!)

**SAMI** Commissioning (see Issue #120) continues to run smoothly, and the Sydney and AAO team hope to make an upgraded SAMI available as a facility instrument in the future.

Remote observing from Epping with **UCLES** has been going well: three nights of service observing have been carried out remotely in the past six months, and another is scheduled for late July. The AAO plans to offer remote observing as an option for experienced UCLES observers in upcoming semesters, and will be trialling remote observing with other AAT instruments as well.

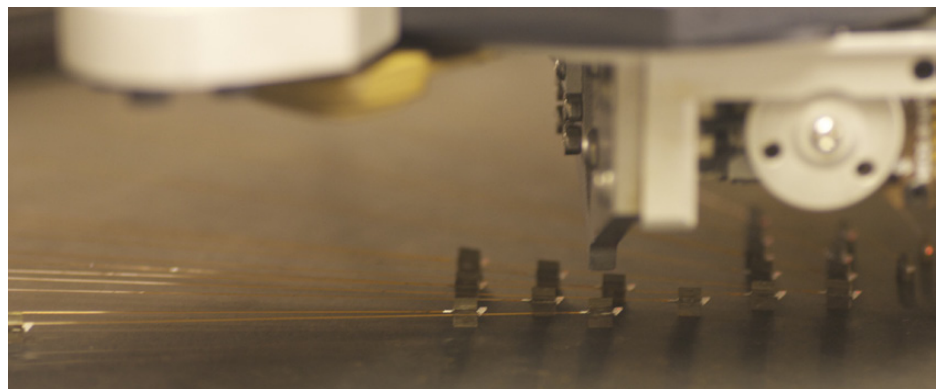

Improvements to the 2dF robotic positioner have greatly reduced field configuration time.

# Letter from Coona

Katrina Harley (AAO)

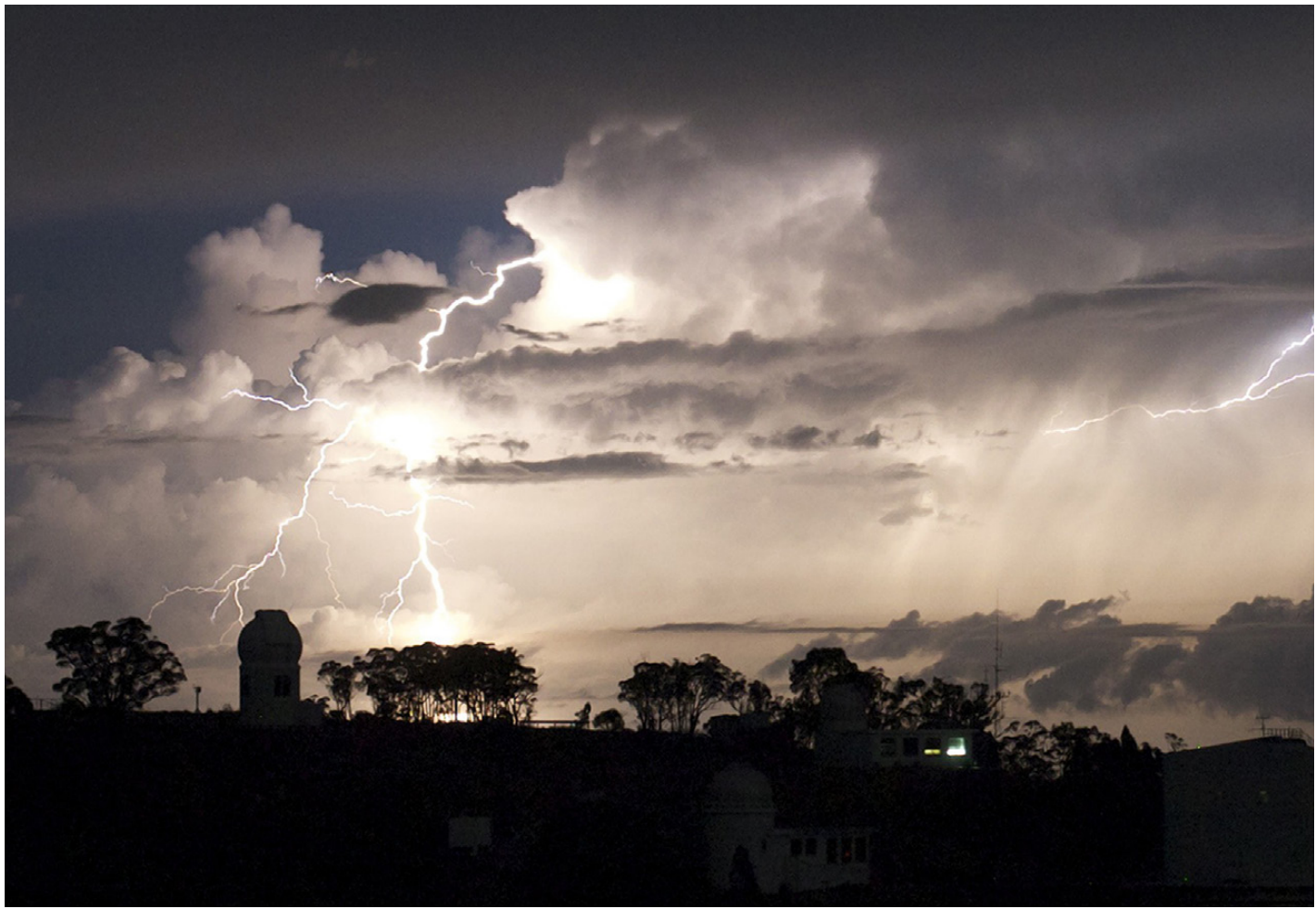

A lightening strike in the distance behind Skymapper from the AAT catwalk. Credit: Andy Green.

Late January and early February has turned out to be very miserable and wet. During this time Coonabarabran has received well over 100 millimetres of rain in just a couple of days, and predicting more rain to come.

In December, Wade Sutherland joined as on-site, fresh from School. Wade is employed as an Electrical Apprentice- Wade is settling in well. Greg Canham has also returned to the AAT to assist in the mechanical workshop while Brendan and Mick are on leave.

Work will commence in the next couple of months to paint the UKST and the AAT domes, it will be a very busy time with many contractors on-site.

The 2011 Christmas party was held at the Acacia Motor Lodge. With drinks and nibbles to start with followed by a very delicious three-course meal, which was enjoyed by all. Many former employees of the AAT attended along with former and current staff from the ANU. It was a fantastic night for all.

A few at site have decided that it's time to get fit, and one way was to participate in the City 2 Surf this year, a few have already started training. We are trying to involve as many people as we can, so if you're interest in getting crushed in a crowd of over 80,000, please nominate yourself.

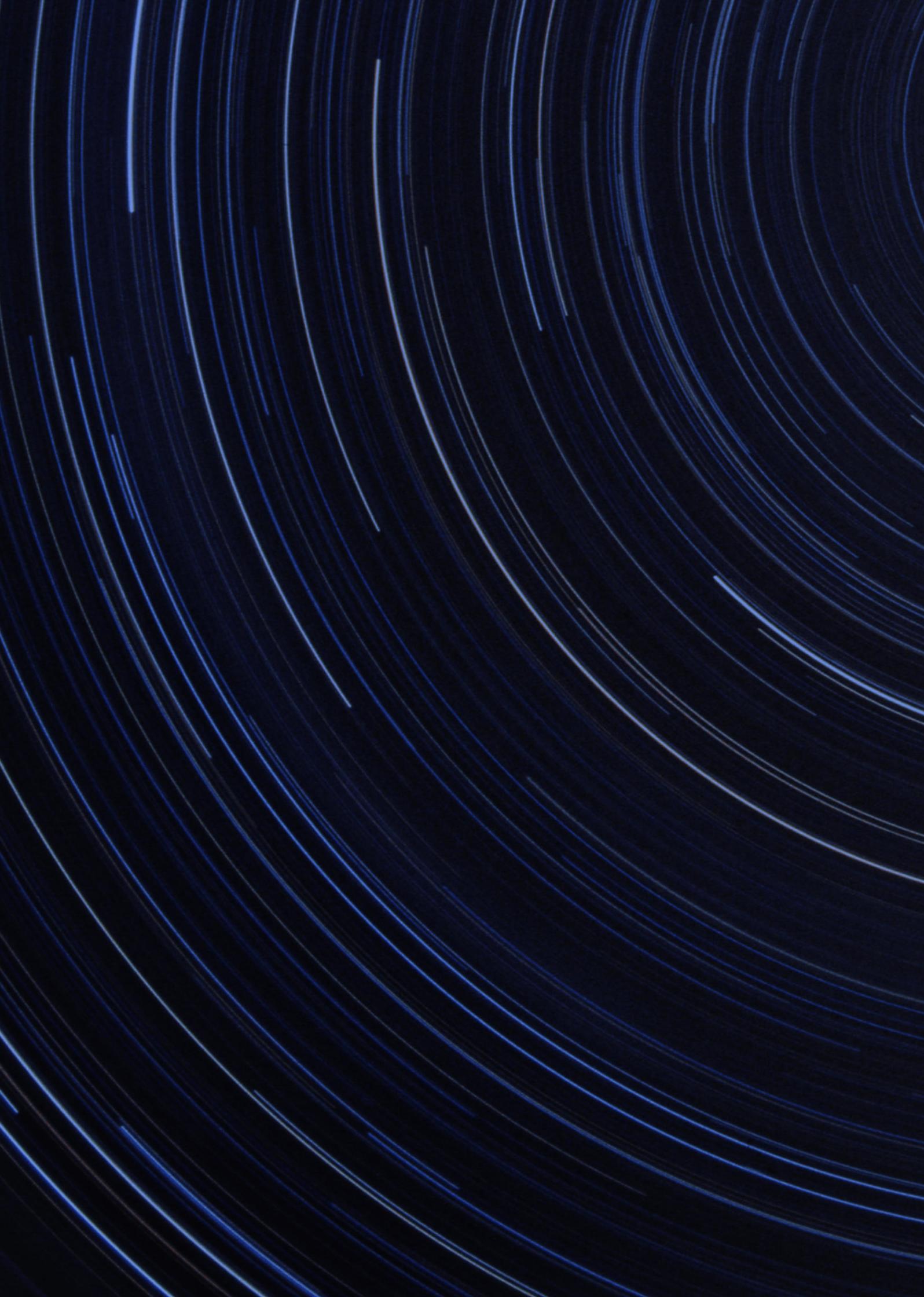

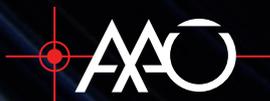